\documentclass[12pt]{iopart}
\usepackage{iopams}
\usepackage{tikz}[thick,scale=1, every node/.style={scale=1.3}]
\usepackage{booktabs}       
\usepackage{physics}
\usepackage{graphicx}
\usepackage{subfig}
\usepackage{float}
\usepackage{amsmath}
\usepackage[colorlinks, citecolor = blue, urlcolor = blue]{hyperref}
\usepackage[backend=biber,style=numeric,autocite=plain,sorting=none]{biblatex}
\DeclareUnicodeCharacter{039B}{\ensuremath{\Lambda}}

\begin{document}
\title[Resonance fluorescence in three-level system]{Resonance fluorescence in $\Lambda$, $V$ and $\Xi$ - type three-level configurations}

\author{Surajit Sen$^{\dag, \ddag, 1}$, Tushar Kanti Dey$^{\ddag, 2}$ and
Bimalendu Deb$^{\S, 3}$}
\address{$^{\dag}$ Department of Physics, Guru Charan College, Silchar 788004, India}
\address{$^\ddag$ Centre of Advanced Studies and Innovation Lab (CASILab)\\ 18/27 Kali Mohan Road, Silchar 788003, India}
\address{$^{\S}$ Indian Association for the Cultivation of Science, Jadavpur, Kolkata 700032, India}
\ead{{\textsuperscript{1}}ssen55@yahoo.com}
\ead{{\textsuperscript{2}}tkdey54@gmail.com}
\ead{{\textsuperscript{3}}msbd@iacs.res.in}
\vspace{10pt}

\begin{abstract}
We theoretically study the resonance fluorescence spectra of the lambda ($\Lambda$), vee ($V$) and cascade ($\Xi$) type three-level configurations. It is shown that each system with two detuning frequencies can be modelled using the $SU(3)$ symmetry group to derive a generalized optical Bloch equation. For each configuration, this equation is solved to calculate the two-time correlation function by invoking the quantum regression theorem. The incoherent part of the power spectra gives the characteristic multi-peak fluorescence profiles which are different for different configurations. We also discuss how the dressed-state structure of such system can explain the origin of quintuplet profile of the fluorescent spectrum. 
\end{abstract}

%
\noindent{\it Keywords}: Resonance Fluorescence, Three-level System, SU(3) Group, Optical Bloch Equation, Power Spectrum 
%
%
%
%

\section{Introduction}
The behavior of two-, three-, or multi-level systems when subjected to an intense electromagnetic field gives rise to numerous intriguing spectral features that may find a wide range of applications in quantum optics and quantum metrology. Historically, a two-level system driven by a strong resonant field was first investigated by Mollow who showed the existence of two symmetrical sidebands in the both sides of the central Rayleigh peak \cite{Mollow1969, Arimondo1996}. In the three-peak spectrum, popularly known as Mollow triplet, the sidebands are located at a shift proportional to the Rabi frequency of the system with their intensity equal to one-third that of the central peak. This phenomenon is known as Resonance Fluorescence (RF) which arises due to the transitions between the dressed states of the effective system formed under the action of an intense external field \cite{Cohen2020}. The predecessors of this light-matter interaction model are, Wigner-Weisskopf model of the spontaneous emission \cite{Weisskopf1930} which leads to Lorentzian spectra, Autler-Townes (AT) effect, where a doublet spectral feature is observed if a two-level system is resonantly driven by a strong oscillatory electromagnetic field \cite{autler1955} and the Fano model \cite{Fano1961} which is characterized by an asymmetric spectrum resulting due to the presence of a continuum state among the states involved. The experimental observation of the anti-bunching of photons in the intensity-intensity correlation is an important outcome of the strongly coupled light-matter quantum dynamics \cite{Kimble1977,Kimble1978}.
\begin{figure}[t!]
\centering
    \subfloat[$\Lambda$ Configuration]{
\resizebox{0.35\textwidth}{!}{%
            \begin{tikzpicture}
\draw[black, thick] (0,9) -- (4,9) node[right] {$ E_3, \ket{3}$};
\draw [black, densely dashed, thick](0,8.8) -- (4.0,8.8);
\draw [stealth-, line width=0.2mm](0.5,9.0) -- (0.5,9.3);
\draw [-stealth, line width=0.2mm](0.5,8.5) -- (0.5,8.8)  node at (-.4,8.9) {$\Delta_{13}^{\Lambda}$} ;
\draw[black, thick] (2.0,6.5) -- (4,6.5) node[right]  {$ E_2,\ket{2}$};
\draw [black, densely dashed, thick](2.0,6.7) -- (4.0,6.7);
\draw [-stealth, line width=0.2mm](2.5,7.0) -- (2.5,6.7) node at (1.6,6.7) {$\Delta_{23}^{\Lambda}$} ;
\draw [stealth-, line width=0.2mm](2.5,6.5) -- (2.5,6.2);
\draw[black, thick] (0,5) -- (4,5)  node[right] {$ E_1,\ket{1}$};
\draw[stealth-stealth, line width=0.25mm] (1,5) -- (1,9)  node[midway,left] {$V_{\pm}$};
\draw[stealth-stealth, line width=0.25mm] (3,6.5) -- (3,9.0) node[midway,right] {$T_{\pm}$};
        \end{tikzpicture}
    }
    }

    \subfloat[$V$ Configuration]{
\resizebox{0.35\textwidth}{!}{%
        \begin{tikzpicture}
\draw[black, thick] (0,9) -- (4,9) node[right] {$ E_3, \ket{3}$};
\draw [black, densely dashed, thick](0,8.8) -- (4.0,8.8);
\draw [stealth-, line width=0.2mm](0.5,9.0) -- (0.5,9.3);
\draw [-stealth, line width=0.2mm](0.5,8.5) -- (0.5,8.8)  node at (-.4,8.9) {$\Delta_{13}^{V}$} ;
\draw[black, densely dashed, thick] (2.0,6.5) -- (4,6.5) node[right]  {$ E_2, \ket{2}$};
\draw [black, thick](2.0,6.7) -- (4.0,6.7);
\draw [-stealth, line width=0.2mm](2.5,7.0) -- (2.5,6.7) node at (1.6,6.7) {$\Delta_{12}^{V}$} ;
\draw [stealth-, line width=0.2mm](2.5,6.5) -- (2.5,6.2);
\draw[black, thick] (0,5) -- (4,5)  node[right] {$ E_1, \ket{1}$};
\draw[stealth-stealth, line width=0.25mm] (1,5) -- (1,9)  node[midway,left] {$V_{\pm}$};
\draw[stealth-stealth, line width=0.25mm] (3,6.7) -- (3,5) node[midway,right] {$U_{\pm}$};
        \end{tikzpicture}
    }      
    }
\subfloat[$\Xi$ Configuration]{
\resizebox{0.35\textwidth}{!}{%
            \begin{tikzpicture}
\draw[black, thick] (0,9) -- (4,9) node[right] {$ E_3, \ket{3}$};
\draw [black, densely dashed, thick](0,8.8) -- (4.0,8.8);
\draw [stealth-, line width=0.2mm](0.5,9.0) -- (0.5,9.3);
\draw [-stealth, line width=0.2mm](0.5,8.5) -- (0.5,8.8)  node at (-.4,8.9) {$\Delta_{23}^{\Xi}$} ;
\draw[black, densely dashed, thick] (0,6.3) -- (4,6.3);
\draw[black, thick] (0,6.5) -- (4,6.5) node[right]  {$ E_2, \ket{2}$};
\draw [-stealth, line width=0.2mm](0.5,6.8) -- (0.5,6.5) node at (-.4,6.4) {$\Delta_{12}^{\Xi}$};
\draw [stealth-, line width=0.2mm](0.5,6.3) -- (0.5,6.0);
\draw[black, thick] (0,5) -- (4,5)  node[right] {$E_1,\ket{3}$};
\draw[stealth-stealth, line width=0.25mm] (1,6.5) -- (1,9)  node[midway,left] {$T_{\pm}$};
\draw[stealth-stealth, line width=0.25mm] (3,6.5) -- (3,5) node[midway,right] {$U_{\pm}$};
        \end{tikzpicture}
   }
    }
\caption{a) The $\Lambda$ configuration is characterized by the transitions $| 1 \rangle$ $\longleftrightarrow$ $| 3 \rangle$ ($V_{\pm}$) and $| 3 \rangle$ $\longleftrightarrow$ $| 2 \rangle$ ($T_{\pm}$) with detuning frequencies $\Delta_{13}^{\Lambda}$ and $\Delta_{23}^{\Lambda}$; b) the $V$ configuration is characterized by the transitions $| 3 \rangle$ $\longleftrightarrow$ $| 1 \rangle$ ($V_{\pm}$) and $|2 \rangle$ $\longleftrightarrow$ $| 1 \rangle$ ($U_{\pm}$) with the detuning frequencies $\Delta_{13}^{V}$ and $\Delta_{12}^{V}$.; c) the $\Xi$ configuration with transitions $| 1 \rangle$ $\longleftrightarrow$ $| 2 \rangle$ ($U_{\pm}$) and $| 2 \rangle$ $\longleftrightarrow$ $| 3 \rangle$ ($T_{\pm}$) with detuning frequencies $\Delta_{23}^{\Xi}$ and $\Delta_{12}^{\Xi}$. The energy $E_i$ ($i=1, 2, 3$) of the uncoupled atomic states $\ket{i}$ are arranged as $E_1 < E_2 < E_3$.}
\end{figure}
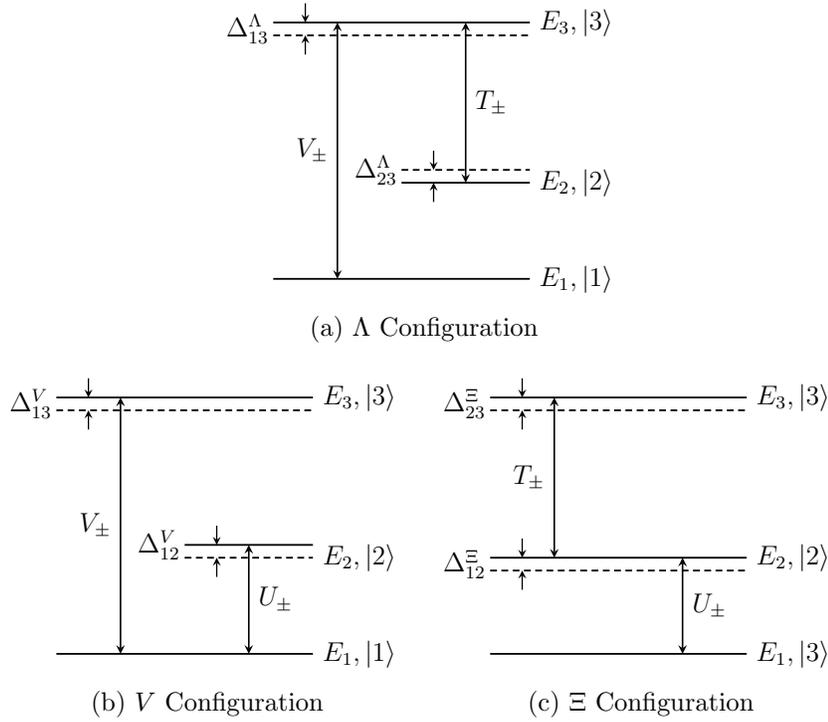
\noindent 
\par
The generalization of the two-level system to the three-level one is quite nontrivial due to three distinct categories of transitions known as, the lambda ($\Lambda$), vee ($V$) and cascade ($\Xi$) type configurations shown in Fig.1. In the recent past the three-level system has drawn considerable attention because it exhibits wide range of quantum-optical phenomena such as two photon coherence \cite{Brewer1975}, double resonance process \cite{Whitley1976}, three-level super-radiance \cite{Bowden1978}, resonance Raman scattering \cite{Sobolewska1976}, population trapping \cite{Arimondo1996}, tri-level echoes \cite{Mossberg1977}, STIRAP \cite{Vitanov2017}, quantum jump \cite{Cook1988}, quantum zeno effect \cite{Misra1977}, three-level Fano effect \cite{Sen2017}, electromagnetically induced transparency (EIT) \cite{Harris1997,Hau1999,Sen2015} etc. In context with the resonance fluorescence, it is revealed that Mollow-like sideband  is not limited to the two-level systems alone, but is also displayed by the multilevel system, including three-level systems driven by two intense driving fields \cite{Narducci1990,Manka1991,Fu1992}. 
\par 
In the eighties, the RF in a two-level system gained prominence due to its inherent significance, particularly with the introduction of damping induced by the squeezed vacuum \cite{Carmichael1987PRA,Carmichael1987}. Since then, the RF in three-level system is being addressed in context with the squeezed vacuum using the $V$ \cite{Jagatap1991,Ferguson1996,Zhou1997} and $\Xi$ type of three-level configurations were studied \cite{Smart1999}. Apart from that, it was further investigated by considering the emission spectrum of two-photon resonant excitation using the $\Xi$ model \cite{Carreno2003}, in the nano-particle system \cite{Carreno2016} and in trapped condition using $\Lambda$ configuration \cite{Bienert2019}. In the realm of three-level configurations, the topology of each configuration is fundamentally different from one another due to their distinct characteristics. Therefore, despite the aforesaid works, it is crucial to adopt a comprehensive approach to study the RF for all three-level configurations particularly to decipher the distinct structure of the dressed states of each system. 

\par
Operationally, a three-level configuration consists of two coupled two-level systems where two dipole transitions are involved shown in Fig.1. Such a system with multiple Lindblad terms takes into account the effects of spontaneous emission and dephasing caused by the vacuum fluctuations in two distinct transition pathways. Consequently, the solution to such a model becomes quite complex to handle theoretically. Recently, the SU(3) group has been shown to have the potential to unveil a wide range of quantum optical phenomena involving three-level systems 
\cite{Yoo1985,Sen2008,Sen2012,Sen2015,Sen2017}. It is therefore interesting to explore the RF of all three-level configurations using this group theoretic method and give a comprehensive comparison of this quantum phenomenon. Our method provides an elegant approach to solve a set of Optical Bloch equations (OBE) and subsequently give the correlation function for fluctuations of the atomic variables around the steady state.
\par
The remaining sections of the paper are organized as follows: In Section II, using the Gell-Mann matrices of $SU(3)$ group, we have revisited the construction of the dissipative $\Lambda$, $V$ and $\Xi$ configurations where the dissipation is taken care of by the Lindblad term. Then in Section-III, the OBE of each configuration is derived. The two-time correlation function using the quantum regression theorem is presented in Section-IV. The power spectra of all configurations are compared and the structure of the dressed states of three-level system is enunciated in Section V to understand the the quintuplet spectrum of resonance fluorescence. In the final Section, we highlight the main results of the paper and discuss the outlook.

\section{The Models}
The dissipative process in the $\Lambda, V$ and $\Xi$ type  three-level configurations can be described by the Liouville equation with Lindblad term ($A=\Lambda, V, \Xi$),
\begin{eqnarray}
\frac{{d\rho ^A }}{{dt}} =  -i\left[ {H^A,\rho^A} \right] + \mathfrak L_D^A,
\label{eq1}
\end{eqnarray}
\noindent
where the density matrix is given by
\begin{eqnarray}
\rho^A=\left(
\begin{array}{ccc}
   {\rho _{33} } & {\rho _{32} } & {\rho _{31} }  \cr
   {\rho _{23} } & {\rho _{22} } & {\rho _{21} }  \cr
   {\rho_{13}} & {\rho _{12} } & {\rho _{11} }  \cr
\end{array}
\right)
\end{eqnarray}
written in the the atomic basis states are $|1\rangle=(0,0,1)^T$, $|2\rangle=(0,1,0)^T$ and $|3\rangle=(1,0,0)^T$ with $\mathfrak L_D^A$ as the Lindblad term. Using Gellmann matrices of the $SU(3)$ representation, the Hamiltonian and Lindblad terms of these configurations are given by \cite{Sen2008,Sen2012},
\numparts 
\begin{eqnarray}\label{eq3a}
H^{\Lambda} =
\left(
\begin{array}{ccc}
\frac{1}{3}\big(\Delta_{13}^{\Lambda}+\Delta_{23}^{\Lambda}\big) & g_{23} & g_{13}\\
g_{23} & \frac{1}{3}\big(\Delta_{13}^{\Lambda}-2\Delta_{23}^{\Lambda}\big) & 0 \\
g_{13} & 0 & -\frac{1}{3}\big(2\Delta_{13}^{\Lambda}-\Delta_{23}^{\Lambda}\big), 
\end{array}
         \right),
\end{eqnarray}
\begin{eqnarray}\label{eq3b}
\mathfrak L_D^\Lambda  &= \Gamma_{31}^\Lambda (V_+\rho^\Lambda V_--\frac{1}{2}V_-V_+\rho^\Lambda-\frac{1}{2}\rho^\Lambda V_-V_+) \nonumber \\ 
&+\Gamma_{32}^\Lambda (T_+\rho^\Lambda T_--\frac{1}{2}T_-T_+\rho^\Lambda-\frac{1}{2}\rho^\Lambda T_-T_+) ,
\end{eqnarray}
\endnumparts
for the $\Lambda$ system with the detuning offsets,  $\Delta_{13}^\Lambda=2\omega_{31}+\omega_{32}-\Omega_{13}$ and  $\Delta_{23}^\Lambda=\omega_{31}+2\omega_{32}-\Omega_{23}$, 
\numparts 
\begin{eqnarray}\label{eq4a}
H^{V} =
\left(
\begin{array}{ccc}
\frac{1}{3}\big(2\Delta_{13}^{V}-\Delta_{12}^{V}\big) & 0 & g_{13}\\
0 & \frac{1}{3}\big(2\Delta_{12}^{V}-\Delta_{13}^{V}\big) & g_{12} \\
g_{13} & g_{12} & -\frac{1}{3}\big(\Delta_{12}^{V}+\Delta_{13}^{V}\big), \\
\end{array}
\right)
\end{eqnarray}
\begin{eqnarray}\label{eq4b}
\mathfrak L_D^V &= \Gamma _{31}^V(V_-\rho^V V_+-\frac{1}{2}V_+V_-\rho^V-\frac{1}{2}\rho^V V_+V_-) \nonumber \\
&+ \Gamma _{21}^V(U_-\rho^V U_+-\frac{1}{2}U_+U_-\rho^V-\frac{1}{2}\rho^V U_+U_-),
\end{eqnarray}
\endnumparts
for the $V$ system  with  $\Delta_{12}^V=\omega_{31}+2\omega_{21}-\Omega_{12}$, $\Delta_{13}^V=2\omega_{31}+\omega_{21}-\Omega_{13}$ and 
\numparts 
\begin{eqnarray}\label{eq5a}
H^{\Xi} =
\left(
\begin{array}{ccc}
\frac{1}{3}\big(\Delta_{12}^{\Xi}+2\Delta_{23}^{\Xi}\big) & g_{23} & 0 \\
g_{23} & \frac{1}{3}\big(\Delta_{12}^{\Xi}-\Delta_{23}^{\Xi}\big) & g_{12} \\
0 & g_{12} & -\frac{1}{3}\big(2\Delta_{12}^{\Xi}+\Delta_{23}^{\Xi}\big) 
\end{array}
         \right),
\end{eqnarray}
\begin{eqnarray}\label{eq5b}
\mathfrak L_D^\Xi &= \Gamma _{32}^\Xi(T_-\rho^\Xi T_+ - \frac{1}{2} T_+T_-\rho^\Xi-\frac{1}{2}\rho^\Xi T_+T_-) \nonumber \\ 
&+ \Gamma _{21}^\Xi(U_-\rho^\Xi U_+ - \frac{1}{2}U_+U_-\rho^\Xi-\frac{1}{2}\rho^\Xi U_+U_-), 
\end{eqnarray}
\endnumparts 
\noindent
for the $\Xi$ system with detuning  $\Delta_{23}^{\Xi}=2\omega_{21}+\omega_{32}-\Omega_{23}$ and  $\Delta_{12}^{\Xi}=2\omega_{32}-2\omega_{21}-\Omega_{12}$. Here, $g_{ij}$ ($i,j=1,2,3$) is the system-field coupling parameter with the laser field with frequency $\Omega_{ij}$, $\Gamma_{ji}$  be the decay constant for the transition pathway $j \rightarrow i$ ($j > i$) and $\omega_{ij}=\omega_{i}-\omega_{j}$ be the difference between transition frequencies of $\ket{i}$-th and $\ket{j}$-th levels with $\hbar \omega_1$ ($=E_1$), $\hbar\omega_2$ ($=E_2$) and $\hbar\omega_3$ ($=E_3$) be the absolute energies of three levels shown in Fig.1. 
The SU(3) shift vectors appearing in Equations (\ref{eq3a}-\ref{eq5b}) are given by \cite{Greiner2012},
\begin{eqnarray}
\eqalign{ 
T_\pm &=\frac{1}{2}(\lambda_1\pm i\lambda_2), \quad
V_\pm =\frac{1}{2}(\lambda_4\pm i\lambda_5), \quad 
U_\pm =\frac{1}{2}(\lambda_6\pm i\lambda_7) \cr 
T_3 &=\lambda_3, \quad V_3=\frac{1}{2}(\sqrt{8}\lambda_8+\lambda_3), \quad
U_3=\frac{1}{2}(\sqrt{8}\lambda_8-\lambda_3).
}
\end{eqnarray}

\noindent
where the Gell-Mann matrices are given by,
\begin{eqnarray}
\eqalign{ 
\lambda_0=&\left(
           \begin{array}{ccc}
             1 & 0 & 0\\
             0 & 1 & 0\\
             0 & 0 & 1\\
           \end{array}
         \right), \quad
\lambda_1=\left(
           \begin{array}{ccc}
             0 & 1 & 0\\
             1 & 0 & 0\\
             0 & 0 & 0\\
           \end{array}
         \right), \quad
\lambda_2=\left(
           \begin{array}{ccc}
             0 & -i & 0\\
             i & 0 & 0\\
             0 & 0 & 0\\
           \end{array}
         \right), \cr
\lambda_3=&\left(
           \begin{array}{ccc}
             1 & 0 & 0\\
             0 & -1 & 0\\
             0 & 0 & 0\\
           \end{array}
         \right), \quad
\lambda_4=\left(
           \begin{array}{ccc}
             0 & 0 & 1\\
             0 & 0 & 0\\
             1 & 0 & 0\\
           \end{array}
         \right), \quad
\lambda_5=\left(
           \begin{array}{ccc}
             0 & 0 & -i\\
             0 & 0 & 0\\
             i & 0 & 0\\
           \end{array}
         \right), \cr
\lambda_6=&\left(
           \begin{array}{ccc}
             0 & 0 & 0\\
             0 & 0 & 1\\
             0 & 1 & 0\\
           \end{array}
         \right), \quad
\lambda_7=\left(
           \begin{array}{ccc}
             0 & 0 & 0\\
             0 & 0 & -i\\
             0 & i & 0\\
           \end{array}
         \right), \quad
\lambda_8=\frac{1}{\sqrt{3}}\left(
           \begin{array}{ccc}
             1 & 0 & 0\\
             0 & 1 & 0\\
             0 & 0 & -2\\
           \end{array}
         \right),
}
\end{eqnarray}
which are normalized as $\lambda_l\lambda_m=\delta_{lm}+d_{lmn}\lambda_n+f_{lmp}\lambda_p$ with $d_{lmn}$ and $f_{lmp}$ ($l,m,n,p=1,2, \dots,8$) being the completely symmetric and completely antisymmetric structure constants. It is worth mentioning here that aforesaid time-independent Hamiltonian appearing in Equations (\ref{eq3a},\ref{eq4a},\ref{eq5a}) can be obtained by using the transformation \cite{Sen2008,Sen2012},
\begin{eqnarray} 
\eqalign{ H^A={U^A}^\dag(t)H^A(t)U^A(t)-i\bigg[\frac{d{U^A}(t)}{dt}\bigg ]{U^A}^{\dag}(t),}
\end{eqnarray}
where the unitary operators are given by,
\numparts 
\begin{eqnarray}
U^{\Lambda}(t) = \exp \big[ - \frac{i}{3}\big((2\Omega _{13}-\Omega _{23})V_3+(2\Omega _{23}-\Omega _{13})T_3\big)\big]t,
\end{eqnarray}
\begin{eqnarray}
U^{V}(t) = \exp \big[- \frac{i}{3}\big((2\Omega _{12}-\Omega _{13})U_3+(\Omega _{13}-2\Omega _{23})V_3 \big)\big]t,
\end{eqnarray}
\begin{eqnarray}
U^{\Xi}(t) = \exp \big[ - \frac{i}{3}\big((2\Omega _{12}+\Omega _{23})U_3+(\Omega _{12}+2\Omega _{23})T_3\big)\big] t,
\end{eqnarray}
\endnumparts
with the time-dependent Hamiltonians of three configurations,  
\numparts 
\begin{eqnarray} 
\fl \qquad H^{\Lambda}(t) = \omega_{31} V_3+\omega_{32} T_3+g_{13} V_+\exp ( - i\Omega _{13} t)+g _{23} T_+\exp(-i\Omega_{23} t)+h.c.,
\end{eqnarray}
\begin{eqnarray} 
\fl \qquad  H^V(t) = \omega_{31}V_3+\omega_{21}U_3+g_{13} V_ + \exp ( - i\Omega _{13} t) + g_{12} U_+\exp(-i\Omega_{12} t)+h.c.,
\end{eqnarray}
\begin{eqnarray} 
\fl \qquad H^\Xi(t) = \omega_{32}T_3+\omega_{21}U_3+g_{12}U_ + \exp(-i\Omega_{12}t)+g_{23}T_+\exp(-i\Omega_{23} t)+h.c..
\end{eqnarray}
\endnumparts
Having knowledge about the model Hamiltonian of all three configurations and the Lindblad term which characterizes the dissipation in such system, we proceed to derive the Optical Bloch Equations. 

\section{Bloch Equation for Three-level Configuration}
The Bloch vector $\mathbf{S}^{A}_{\mathbb{P}_i}(t)$ of a generic three-level configuration is defined as,
\begin{eqnarray}
{\mathbf{S}}^{A}_{\mathbb{P}_i}(t) =Tr[\rho^A(t){\mathbb{P}_i}], \label{eq11}
\end{eqnarray}
where $\mathbb{P}_i$ ($T_{+,-,3},V_{+,-,3},U_{+,-,3}$) is the $SU(3)$ shift operators. Using the algebra of the shift vectors it is easy to see that,
\begin{eqnarray}\label{eq12}
\frac{d\mathbf{S}^A_{\mathbb{P}_i}(t)}{dt} = Tr\big[\frac{d\rho^A(t)}{dt}{\mathbb{P}_i}\big], \cr 
=-i\big\langle\big[H^A,\rho^A\big]{\mathbb{P}_i}\big\rangle+ \big \langle \mathfrak L_D^A{\mathbb{P}_i}\big \rangle,
\end{eqnarray}
where Equation (\ref{eq1}) is used along with $\big \langle {\mathcal{\hat{O}}} \big \rangle^A = \Tr[{\mathcal{\hat{O}}} \rho^A]$. Using the Lindblad terms in Equations (\ref{eq3b}, \ref{eq4b}, \ref{eq5b}), Equation (\ref{eq12}) for three configurations are given by, 
\numparts 
\begin{eqnarray}\label{eq13a}
\fl \frac{d\mathbf{S}^\Lambda_{\mathbb{P}_i}(t)}{dt} = -i\Tr\big[(H^\Lambda\rho^\Lambda-\rho^\Lambda H^\Lambda){\mathbb{P}_i}\big]
+\Gamma_{31}Tr\big[(V_+\rho^\Lambda V_- - \frac{1}{2}\rho^\Lambda V_-V_+ -\frac{1}{2}V_-V_+ \rho^\Lambda){\mathbb{P}_i}\big] \nonumber \\
+\Gamma_{32}Tr\big[(T_+\rho^\Lambda T_- -\frac{1}{2}\rho^\Lambda T_-T_+ -\frac{1}{2}T_-T_+\rho^\Lambda){\mathbb{P}_i}\big],
\end{eqnarray}
\begin{eqnarray}\label{eq13b}
\fl \frac{d\mathbf{S}^V_{\mathbb{P}_i}(t)}{dt} = -i\Tr\big[(H^V\rho^V-\rho^V H^V){\mathbb{P}_i}\big] +\Gamma_{31}Tr\big[(V_-\rho^\Lambda V_+-\frac{1}{2}\rho^\Lambda V_+V_- -\frac{1}{2}V_+V_-\rho^\Lambda\big){\mathbb{P}_i}] \nonumber \\
+\Gamma_{21}Tr\big[(U_-\rho^V U_+ - \frac{1}{2}\rho^V U_+U_-  - \frac{1}{2}U_+U_-\rho^V){\mathbb{P}_i}\big],
\end{eqnarray}
\begin{eqnarray}\label{eq13c}
\fl \frac{d\mathbf{S}^{\Xi}_{\mathbb{P}_i}(t)}{dt} = -i\Tr\big[(H^\Xi\rho^\Xi-\rho^\Xi H^\Xi){\mathbb{P}_i}\big]+\Gamma_{21}Tr\big[(U_-\rho^\Xi U_+ - \frac{1}{2}\rho^\Xi U_+U_- - \frac{1}{2}U_+U_-\rho^\Xi){\mathbb{P}_i}\big]  \nonumber \\
+\Gamma_{32}Tr\big[(T_- \rho^\Xi T_+- \frac{1}{2}\rho^\Xi T_+T_- - \frac{1}{2}T_+ T_-\rho^\Xi){\mathbb{P}_i}\big],
\end{eqnarray}
\endnumparts 
Thus we note that only a pair of SU(3) projection vectors are involved for a specific configuration. 

To derive the OBE of the $\Lambda$ configuration, we first substitute the shift vector $\mathbb{P}_i =V_{\pm},V_3,T_{\pm},T_3,U_{\pm},U_3$ in Equation (\ref{eq13a}) and obtain,
\numparts 
\begin{eqnarray}\label{eq14a}
\fl \frac{dS_{T_{+}}^\Lambda(t)}{dt} = -i\Tr\big[(H^\Lambda\rho^\Lambda-\rho^\Lambda H^\Lambda)T_+\big]
+\Gamma_{31}Tr\big[(V_+\rho^\Lambda V_- - \frac{1}{2}\rho^\Lambda V_- V_+ -\frac{1}{2}V_-V_+ \rho^\Lambda)T_+\big] \nonumber \\
+\Gamma_{32}Tr\big[(T_+\rho^\Lambda T_- -\frac{1}{2}\rho^\Lambda T_-T_+ - \frac{1}{2}T_- T_+ \rho^\Lambda)T_+\big],
\end{eqnarray}
\begin{eqnarray}
\fl \frac{dS_{T_{-}}^\Lambda(t)}{dt} = -i\Tr\big[(H^\Lambda\rho^\Lambda-\rho^\Lambda H^\Lambda)T_{-}\big]
+\Gamma_{31}Tr\big[(V_+ \rho^\Lambda V_- - \frac{1}{2}\rho^\Lambda V_- V_+ -\frac{1}{2}V_-V_+ \rho^\Lambda) T_-\big] \nonumber \\
+\Gamma_{32}Tr\big[T_+ \rho^\Lambda T_-  - \frac{1}{2}\rho^\Lambda T_-T_+ - \frac{1}{2}T_-T_+  \rho^\Lambda)T_-\big],
\end{eqnarray}
\begin{eqnarray}
\fl \frac{dS_{T_{3}}^\Lambda(t)}{dt} = -i\Tr\big[(H^\Lambda\rho^\Lambda-\rho^\Lambda H^\Lambda)T_{3}\big]
+\Gamma_{31}Tr\big[(V_+ \rho^\Lambda V_- -  \frac{1}{2}\rho^\Lambda V_- V_+ -\frac{1}{2}V_-V_+\rho^\Lambda\big)T_3\big] \nonumber \\
+\Gamma_{32}Tr\big[(T_+\rho^\Lambda T_- -\frac{1}{2}T_-T_+\rho^\Lambda - \frac{1}{2}\rho^\Lambda T_-T_+) T_3\big],
\end{eqnarray}
\begin{eqnarray}
\fl \frac{dS_{V_{+}}^\Lambda(t)}{dt} = -i\Tr\big[(H^\Lambda\rho^\Lambda-\rho^\Lambda H^\Lambda)V_+\big]
+\Gamma_{31}Tr\big[(V_+\rho^\Lambda V_- -\frac{1}{2}\rho^\Lambda V_-V_+ - \frac{1}{2}V_- V_+ \rho^\Lambda)V_+\big] \nonumber \\
+\Gamma_{32}Tr\big[T_+\rho^\Lambda T_- -\frac{1}{2}\rho^\Lambda T_-T_+ - \frac{1}{2}T_- T_+ \rho^\Lambda)V_+\big],
\end{eqnarray}
\begin{eqnarray}
\fl \frac{dS_{V_{-}}^\Lambda(t)}{dt} = -i\Tr\big[(H^\Lambda\rho^\Lambda-\rho^\Lambda H^\Lambda)V_{-}\big]
+\Gamma_{31}Tr\big[(V_+\rho^\Lambda V_- -\frac{1}{2} \rho^\Lambda V_-V_+ - \frac{1}{2} V_-V_+  \rho^\Lambda)V_-\big] \nonumber \\
+\Gamma_{32}Tr\big[(T_+\rho^\Lambda T_- -\frac{1}{2} \rho^\Lambda T_-T_+ - \frac{1}{2}T_- T_+ \rho^\Lambda)V_-\big],
\end{eqnarray}
\begin{eqnarray}
\fl \frac{dS_{V_{3}}^\Lambda(t)}{dt} = -i\Tr\big[(H^\Lambda\rho^\Lambda-\rho^\Lambda H^\Lambda)V_{3}\big]
+\Gamma_{31}Tr\big[(V_+\rho^\Lambda V_- -\frac{1}{2} \rho^\Lambda V_-V_+ - \frac{1}{2}V_-V_+ \rho^\Lambda)V_3\big] \nonumber \\
+\Gamma_{32}Tr\big[(T_+\rho^\Lambda T_- -\frac{1}{2} \rho^\Lambda T_-T_+ - \frac{1}{2}T_-T_+\rho^\Lambda) V_3\big],
\end{eqnarray}
\begin{eqnarray}
\fl \frac{dS_{U_{+}}^\Lambda(t)}{dt} = -i\Tr\big[(H^\Lambda\rho^\Lambda-\rho^\Lambda  H^\Lambda)U_{+}\big]
+\Gamma_{31}Tr\big[(V_+\rho^\Lambda V_- -\frac{1}{2} \rho^\Lambda V_-V_+ - \frac{1}{2}V_- V_+  \rho^\Lambda )U_+\big]  \nonumber \\
+\Gamma_{32}Tr\big(T_+ \rho^\Lambda T_- -\frac{1}{2} \rho^\Lambda T_-T_+ - \frac{1}{2}T_- T_+  \rho^\Lambda) U_+\big],
\end{eqnarray}
\begin{eqnarray}
\fl \frac{dS_{U_{-}}^\Lambda(t)}{dt} = -i\Tr\big[(H^\Lambda\rho^\Lambda-\rho^\Lambda H^\Lambda)U_{-}\big]
+\Gamma_{31}Tr\big[(V_+\rho^\Lambda V_- -\frac{1}{2} \rho^\Lambda V_-V_+ - \frac{1}{2}V_-V_+   \rho^\Lambda)U_-\big] \nonumber \\
+\Gamma_{32}Tr\big[(T_+\rho^\Lambda T_- -\frac{1}{2} \rho^\Lambda T_-T_+ - \frac{1}{2}T_- T_+ \rho^\Lambda)U_-\big],
\end{eqnarray}
\begin{eqnarray}\label{eq14i}
\fl \frac{dS_{U_{3}}^\Lambda(t)}{dt} = -i\Tr\big[(H^\Lambda\rho^\Lambda-\rho^\Lambda  H^\Lambda)U_{3}\big]
+\Gamma_{31}Tr\big[(V_+ \rho^\Lambda V_- -\frac{1}{2} \rho^\Lambda V_+V_- - \frac{1}{2}V_+V_-  \rho^\Lambda )U_3\big]  \nonumber \\
+\Gamma_{32}Tr\big[(T_+\rho^\Lambda T_- -\frac{1}{2} \rho^\Lambda T_+T_- - \frac{1}{2}T_+T_- \rho^\Lambda)U_3\big],
\end{eqnarray}
\endnumparts 
\noindent
Equation (\ref{eq11}) supplemented by the normalization condition $\Tr\rho^{\Lambda}(t)=1$ gives the elements of the density matrix in terms of the Bloch vectors, 
\begin{equation}\label{eq15}
\eqalign{\rho_{33}^\Lambda&=\frac{1}{3}(1+S_{T_{3}}^\Lambda+S_{V_{3}}^\Lambda), \quad \rho_{32}^\Lambda=S_{T_{-}^\Lambda}, \quad \rho_{31}^\Lambda=S_{V_{-}}^\Lambda, \\
\rho_{23}^\Lambda&=S_{T_{+}}^\Lambda, \quad \rho_{22}^\Lambda=\frac{1}{3}(1-S_{T_{3}}^\Lambda+S_{U_{3}}^\Lambda),
\quad \rho_{21}=S_{U_{-}}^\Lambda, \\
\rho_{13}^\Lambda&=S_{V_{+}}^\Lambda, \quad \rho_{12}^\Lambda=S_{U_{+}}^\Lambda, \quad \rho_{11}^\Lambda=\frac{1}{3}(1-S_{U_{3}}^\Lambda-S_{V_{3}}^\Lambda).} 
\end{equation}
Finally substituting the elements of the density matrix from Equation ({\ref{eq15}}) into (\ref{eq14a}-\ref{eq14i}), we obtain the desired OBE of the $\Lambda$ configuration,
\begin{eqnarray}\label{eq16}
\frac{d\mathbf{S^\Lambda_{\mathbb{P}_i}}(t)}{dt}= M^\Lambda\mathbf{S^\Lambda_{\mathbb{P}_i}}+\mathbf{B^\Lambda},
\end{eqnarray}
where the Bloch matrix $M^{\Lambda}$ and the inhomogeneous term  $\mathbf{B^\Lambda}$ are given in Appendix. Proceeding similar way, the OBE for the $V$ and $\Xi$ type configurations are obtained as shown in Appendix.

\section{Power Spectrum}
The evaluation of the power spectrum requires the information about the two-time correlation function of the fluctuation about the steady state. This is usually done by using the quantum regression theorem. In the long time limit (${t\rightarrow \infty}$) when the system attains the steady state we have, 
\begin{eqnarray}
\frac{d\langle\mathbf{S^{A}_{\mathbb{P}_i}(t)\rangle}}{dt}\bigg|_{t\rightarrow \infty} = 0 = M^\Lambda \mathbf{\langle S_{{\mathbb{P}_i}}^A(\infty)\rangle}_{s}+\mathbf{B^A}.
\end{eqnarray}
Now introducing the fluctuation of the atomic variable around the steady state, namely, $\langle\delta{\mathbf{S^{A}_{\mathbb{P}_i}}(t)}\rangle=\mathbf{S^{A}_{\mathbb{P}_i}}(t)-\langle \mathbf{S^{A}_{\mathbb{P}_i}(\infty)\rangle}_{s} \equiv \langle \langle \mathbb{P}_i(t)\rangle\rangle$, the in-homogeneous term $\mathbf{B^A}$ can be eliminated,  
\begin{eqnarray}\label{eq18}
\frac{d\langle \langle \mathbb{P}_i(t)\rangle\rangle}{dt}= M^A_{} \langle \langle \mathbb{P}_i(t)\rangle\rangle.
\end{eqnarray}
According to the quantum regression theorem \cite{Walls2007}, the two-time correlation function $\mathbf{K}^{A}_{\mathbb{P}_i\mathbb{P}_j}(\tau)$ is given by 
\begin{eqnarray}
\frac{d\mathbf{K}^{A}_{\mathbb{P}_i\mathbb{P}_j}(\tau)}{d\tau}= M^A \mathbf{K}^{A}_{\mathbb{P}_i\mathbb{P}_j}(\tau),
\end{eqnarray}
where the incoherent part of the correlation function is given by,
\begin{eqnarray}
\mathbf{K}^{\Lambda}_{\mathbb{P}_i\mathbb{P}_j}(\tau)=
\lim_{t \rightarrow \infty}\langle {\delta\mathbf{ S^{\Lambda}_{\mathbb{P}_i}}}(t+\tau) \mathbf{\delta S^{\Lambda}_{\mathbb{P}_j}(t)} \rangle
\equiv \langle \langle \mathbb{P}_i(\tau)\mathbb{P}_j(0)\rangle\rangle. 
\end{eqnarray}
with the two-time correlation function depends upon the difference of time $\tau$.
\par
In the case of the $\Lambda$ configuration ($A=\Lambda$), in order to calculate the power spectrum for the $\langle\langle V_{+}(\tau)V_{-}(0)\rangle\rangle$  in transition pathway ${3} \rightarrow {1}$, we need to solve the equation, 
\numparts 
\begin{eqnarray}\label{eq21a}
\frac{d\mathbf{K}^{\Lambda}_{V_+\mathbb{P}_j}(\tau)}{d\tau}= M^{\Lambda}
\mathbf{K}^{\Lambda}_{V_+\mathbb{P}_j}(\tau), 
\end{eqnarray}
where the correlation vector $\mathbf{K}^{\Lambda}_{V_+\mathbb{P}_j}(\tau)$ is given by,
\begin{eqnarray}
\mathbf{K}^{\Lambda}_{V_+\mathbb{P}_j}(\tau)=&\big(\langle\langle V_+(\tau)T_+(0) \rangle\rangle,  \langle\langle V_+(\tau)T_-(0) \rangle\rangle, \langle\langle V_+(\tau)T_3(0) \rangle\rangle,  \nonumber \\ 
&  \langle\langle V_+(\tau)V_+(0) \rangle\rangle,  \langle\langle V_+(t)V_-(0) \rangle\rangle,  \langle\langle V_+(\tau)V_3(0) \rangle\rangle, \\ 
&  \langle\langle V_+(\tau)U_+(0) \rangle\rangle,  \langle\langle V_+(\tau)U_-(0) \rangle\rangle,  \langle\langle V_+(\tau)U_3(0) \rangle\rangle \big)^T,  \nonumber
\end{eqnarray}
To evaluate the correlation function, we first obtain the initial conditions using GellMann algebra,
\begin{eqnarray}\label{eq21c}
K_{V_{+}T_{+}}^{A}(0) = 0, K_{V_{+}T_{-}}^{A}(0) = 0, K_{V_{+}T_{3}}^{A}(0) = 0,  
K_{V_{+}V_{+}}^{A}(0) = 0, \nonumber \\
K_{V_{+}V_{-}}^{\Lambda}(0) = 
\frac{1}{3}(1 + \langle \langle {T_3}\rangle\rangle_s + \langle\langle {V_3}\rangle\rangle_s), \quad 
K_{V_{+}V_{3}}^{\Lambda}(0) = -\langle\langle {V_+}\rangle\rangle_s, \\
K_{V_{+}U_{+}}^{\Lambda}(0) =0, \quad 
K_{V_{+}U_{-}}^{\Lambda}(0) = \langle\langle {T_+}\rangle\rangle_s, \quad 
K_{V_{+}U_{3}}^{\Lambda}(0) = -\langle\langle {V_+}\rangle \rangle_s \nonumber 
\end{eqnarray}
\endnumparts
where $\langle \langle \mathbb{P}_i\rangle\rangle_s$ is obtained from Equation (\ref{eq18}) under the steady state condition.
\par 
Proceeding in a similar way, to find the correlation vector 
$\mathbf{K}^{\Lambda}_{T_+\mathbb{P}_j}(\tau)$ and hence $\langle\langle T_{+}(\tau)T_{-}(0)\rangle \rangle$, we solve,
\numparts 
\begin{eqnarray}\label{eq22a}
\frac{d\mathbf{K}^{\Lambda}_{T_+\mathbb{P}_j}(\tau)}{d\tau}= M^{\Lambda}
\mathbf{K}^{\Lambda}_{T_+\mathbb{P}_j}(\tau),
\end{eqnarray}
where the correlation vector is given by
\begin{eqnarray}
\mathbf{K}^{\Lambda}_{T_+\mathbb{P}_j}(\tau)=&\big( \langle\langle T_+(\tau)T_+(0) \rangle, \langle\langle T_+(\tau)T_-(0) \rangle\rangle, \langle\langle T_+(\tau)T_3(0) \rangle\rangle,  \nonumber \\ 
& \langle\langle T_+(\tau)V_+(0) \rangle, \langle\langle T_+(\tau)V_-(0) \rangle\rangle, \langle\langle T_+(\tau)V_3(0) \rangle\rangle, \\
& \langle\langle T_+(\tau)U_+(0) \rangle\rangle, \langle\langle T_+(\tau)U_-(0) \rangle\rangle, \langle\langle T_+(\tau)U_3(0) \rangle\rangle 
\big)^T, \nonumber
\end{eqnarray}
with the initial conditions,
\begin{eqnarray}\label{eq22c}
K_{T_{+}T_{+}}^{A}(0) = 0, 
K_{T_{+}T_{-}}^{A}(0) = \frac{1}{3}(1 + \langle\langle {T_3}\rangle\rangle_{s} + \langle\langle {V_3}\rangle\rangle_{s}),  \nonumber \\
K_{T_{+}T_{3}}^{A}(0) = - \langle\langle {T_+}\rangle\rangle_{s}, 
K_{T_{+}V_{+}}^{A}(0) = 0, 
K_{T_{+}V_{-}}^{\Lambda}(0) = 0, 
K_{T_{+}V_{3}}^{\Lambda}(0) = 0, \\
K_{T_{+}U_{+}}^{\Lambda}(0) = \langle\langle {V_+}\rangle\rangle_{s}, \quad 
K_{T_{+}U_{-}}^{\Lambda}(0) = 0, \quad 
K_{T_{+}U_{3}}^{\Lambda}(0) = \langle\langle {T_+}\rangle\rangle_{s} \nonumber
\end{eqnarray}
\endnumparts
\par
Finally having knowledge about the correlation function, we proceed to evaluate the power spectrum of the three-level configuration given by  \cite{Ferguson1996,Meystre2007}, 
\begin{eqnarray}\label{eq23}
S_{ij}^{A}(\omega)=Re\bigg\{\int_0^\infty d\tau \mathbf{K}^{A}_{P_iP_j}(\tau) e^{-i(\omega-\Omega_{jk})\tau}\bigg\}.
\end{eqnarray}
where $\omega$ be the frequency of the fluorescent light emanating from the three-level system corresponding to the transition $\ket{i} \leftrightarrow \ket{j}$. To compute the incoherent spectrum, we solve the evolution equations of the correlation function $\mathbf{K}^{A}_{P_iP_j}(\tau)$ using Laplace transformation with the initial conditions discussed above. Taking two detuning offsets at resonant frequencies, $\Delta_{ij}^{A}\approx 0$ and $\Delta_{jk}^{A}\approx 0$, the incoherent spectrum can be given by the algabraic equation \cite{Meystre2007,Ferguson1996}, 
\begin{eqnarray}
S_{ij}^{A}(\omega) \approx Re\bigg\{
\mathbf{K}^{A}_{P_iP_j}(s)\bigg|_{s=-i(\omega-\Omega_{ij}^A)}^{\Delta_{ij}=0, \Delta_{jk}=0}\bigg\}.
\end{eqnarray}
with the Laplacian parameter given by $s \approx -i(\omega-\Omega_{ij}^A)$. 
\par 
For example, in the $\Lambda$ configuration, we have two distinct transition pathways $|1\rangle \leftrightarrow |3\rangle$ and $|2\rangle \leftrightarrow |3\rangle$ which correspond to the correlation function $\langle \langle V_{+}(\tau)V_{-}(0)\rangle\rangle$ and $\langle\langle T_{+}(\tau)T_{-}(0)\rangle\rangle$. Then for the  pathway $|1\rangle \leftrightarrow |3\rangle$, the power spectrum is given by, 
\numparts 
\begin{eqnarray}
S_{13}^{\Lambda}(\omega) \approx Re\bigg\{
\mathbf{K}^{\Lambda}_{V_+V_-}(s)\bigg|_{s=-i(\omega-\Omega_{13}^\Lambda)}^{\Delta_{13}=0,\Delta_{23}=0}\bigg\},
\end{eqnarray}
and for the transition $\ket{2} \leftrightarrow \ket{3}$ we obtain,
\begin{eqnarray}
S_{23}^{\Lambda}(\omega) \approx Re\bigg\{
\mathbf{K}^{\Lambda}_{T_+T_-}(s)\bigg|_{s=-i(\omega-\Omega_{23}^\Lambda)}^{\Delta_{13}=0,\Delta_{23}=0}\bigg\},
\end{eqnarray}
\endnumparts
respectively. 
\par 
The evaluation of the power spectrum for the $V$ and $\Xi$ configurations is similar where the pairs of correlation functions, $\{\expval{\expval{V_{+}(\tau)V_{-}(0)}}, \expval{\expval{U_{+}(\tau)U_{-}(0)}}\}$ and \\ $\{\expval{\expval{U_{+}(\tau)U_{-}(0)}}, \expval{\expval{T_{+}(\tau)T_{-}(0)}}\}$ are relevant. 

\section{Results and Discussions}
We now turn our attention to the resonance fluorescence profiles of different configurations. To get the desired power spectrum of the $\Lambda$ configuration for two distinct transition pathways, we have to solve Equations (\ref{eq21a}) and (\ref{eq22a}), subject to the steady state conditions given by Equations (\ref{eq21c}) and (\ref{eq22c}). 
It is convenient to work with the dimensionless parameters, $\tilde{\Delta}_{ij}=\Delta_{ij}/\Gamma$, $\tilde{\Gamma}_{ij}=\Gamma_{ij}/\Gamma$, $\tilde{g}_{ij}=g_{ij}/\Gamma$, where the scaling factor $\Gamma$ is different for different configurations (See figures 2-7). As mentioned earlier, we solve the correlation function using Laplace transformation method while taking the resonant detuning to be $\Delta_{13}^{\Lambda} \approx 0$ and $\Delta_{23}^{\Lambda} \approx 0$ and then, retrace back to frequency domain with the Laplace variables $s =  -i (\omega - \Omega_{13})$ or $s =-i(\omega-\Omega_{23})$ \cite{Ferguson1996,Meystre2007}. 
\par 
 The power spectra of $\Lambda$ configuration are plotted in Fig.2 and 3 for various combination of input parameters. It is noteworthy that in the strong coupling regime with $g_{23} >1$ and $g_{13}>1$, we observe distinctive quintuplet spectrum as depicted in Fig.2(a,b) and Fig.3(a,b) for the transition pathways, $\ket{3} \rightarrow \ket{1}$ and $\ket{3} \rightarrow \ket{2}$, respectively. In contrast, when the laser coupling in one transition is weak and that in the other transition is strong, then we 
 observe Mollow-like triplet in the RF of the latter transition as shown in Figs. 2(d) and 3(c), while a doublet in the RF profile depicted in Figs.2(c) and 3(d). The appearance of a Mollow-like triplet structure clearly indicates that the system effectively reduces to a strongly driven two-level atom when the laser-coupling to the other transition pathway for which the RF is not observed goes to zero. The emergence of a doublet structure is a signature of the RF counterpart where we have familiar Autler-Townes splitting. A strong laser-coupling in $\ket{3} \rightarrow \ket{2}$ ($\ket{3} \rightarrow \ket{1}$) transition creates two excited dressed states the signature of which is then manifested as a doublet in the RF in weakly coupled $\ket{3} \rightarrow \ket{1}$ ($\ket{3} \rightarrow \ket{2}$) transition.
\par 
In a similar way, we can proceed to discuss the spectral features of the $V$-type configuration as shown in Figs. 4 and 5 which exhibit distinct shapes of the fluorescent spectra. The distinct shapes of the fluorescent spectra contrast to the $\Lambda$ configurations is clearly evident. The RF profiles of the $\Xi$ -type configuration are displayed in Fig.6 and 7. Once again, with strong couplings for both transition pathways we observe the quintet resonance profile. However, in the weak coupling regime for one transition and strong-coupling for other one, we observe a Mollow-like triplet or a doublet profile only for the transition $\ket{2} \leftrightarrow \ket{1}$. In contrast, for the $\ket{3} \leftrightarrow \ket{2}$ transition, we witness five-peak spectra with different shapes. 
\par 
At this point, it is important to emphasize that the magnitude of some specific profiles in the $V$ and $\Xi$ configurations which are quite small compare to other profiles as revealed in Fig.4(a,c), 5(b,d) for the $V$ configuration and Fig.6c, 7(c,d) for the $\Xi$ configuration. In essence, the unique spectral signatures of the resonance fluorescence serve as another key identifying feature to differentiate each three-level configuration from the other two. Among various profiles, appearance of the quintuplet resonance profile for different type of three-level configurations is in agreement with the results of previous theoretical investigations \cite{Narducci1990,,Manka1991,Fu1992,Ferguson1996}, however, in different parameter regimes, the emergence of the Mollow-like triplet and Autler-Townes-like doublet is noteworthy. 
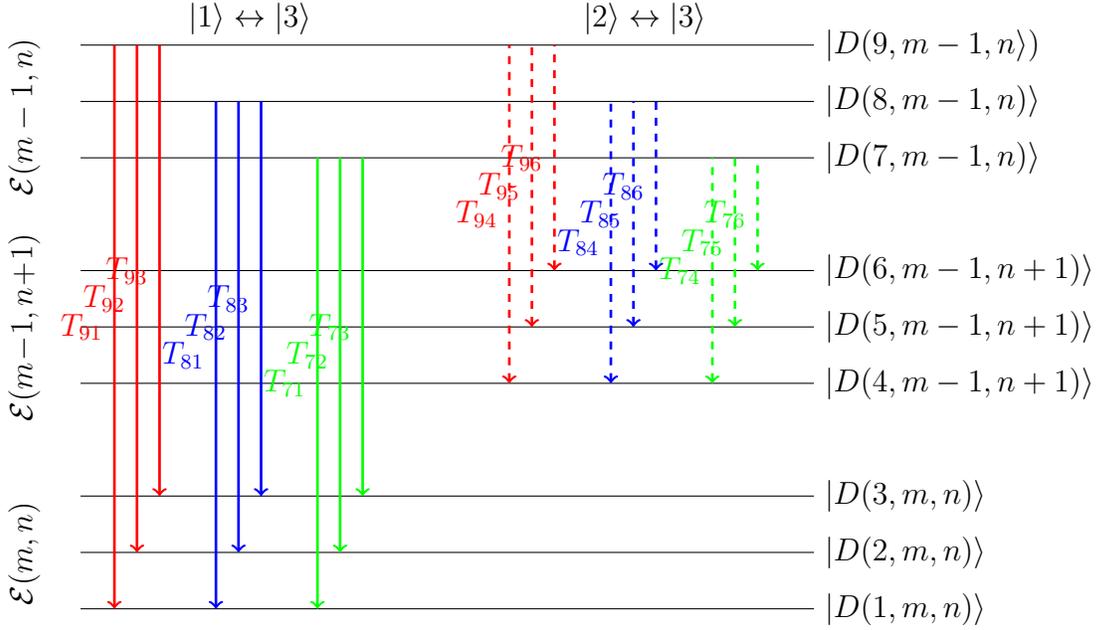
\begin{figure}[ht]
\centering  
\begin{tikzpicture}[scale=1.5]
\draw (0,5.0) node[left] {${}$} -- (6.5,5.0) node[right] {$\ket{D(9,m-1, n})$};
\draw (0,4.5) node[left] {} -- (6.5,4.5) node[right] {$\ket{D(8,m-1,n)}$};
\draw (0,4.0) node[left] {${}$} -- (6.5,4.0) node[right] {$\ket{D(7,m-1,n)}$};
\draw (0,3.0) node[left] {${}$} -- (6.5,3.0) node[right] {$\ket{D(6,m-1, n+1)}$};
\draw (0,2.5) node[left] {} -- (6.5,2.5) node[right] {$\ket{D(5,m-1,n+1)}$};
\draw (0,2.0) node[left] {${}$} -- (6.5,2.0) node[right] {$\ket{D(4,m-1,n+1)}$};
\draw (0,1.0) node[left] {${}$} -- (6.5,1.0) node[right] {$\ket{D(3,m,n)}$};
\draw (0,0.5) node[left] {${}$} -- (6.5,.5) node[right] {$\ket{D(2,m,n)}$};
\draw (0,0) node[left] {${}$} -- (6.5,0) node[right] {$\ket{D(1,m,n)}$};
  \draw[<-, red, line width=1pt] (0.3,0) -- (0.3,5) node[midway, left] {$T_{91}$};
  \draw[<-, red, line width=1pt] (0.5,0.5) -- (0.5,5)node[midway, left] {$T_{92}$};
  \draw[<-, red, line width=1pt] (0.7,1.0) -- (0.7,5)node[midway, left] {$T_{93}$};
  \draw[<-, blue, line width=1pt] (1.2,0.0) -- (1.2,4.5)node[midway, left] {$T_{81}$};
  \draw[<-, blue, line width=1pt] (1.4,0.5) -- (1.4,4.5) node[midway, left] {$T_{82}$};
  \draw[<-, blue, line width=1pt] (1.6,1.0) -- (1.6,4.5)node[midway, left] {$T_{83}$};
  \draw[<-, green, line width=1pt] (2.1,0.0) -- (2.1,4.0) node[midway, left] {$T_{71}$};
  \draw[<-, green,line width=1pt] (2.3,0.5) -- (2.3,4.0)node[midway, left] {$T_{72}$};
  \draw[<-, green, line width=1pt] (2.5,1.0) -- (2.5,4.0)node[midway, left] {$T_{73}$};
  \draw[<-, red, dashed, line width=1pt] (3.8,2.0) -- (3.8,5) node[midway, left] {$T_{94}$};
  \draw[<-, red, dashed,, line width=1pt] (4.0,2.5) -- (4.0,5) node[midway, left] {$T_{95}$};
  \draw[<-, red, dashed,, line width=1pt] (4.2,3.0) -- (4.2,5) node[midway, left] {$T_{96}$};
\draw[<-, blue, dashed, line width=1pt] (4.7,2.0) -- (4.7,4.5) node[midway, left] {$T_{84}$};
\draw[<-,  blue, dashed, line width=1pt] (4.9,2.5) -- (4.9,4.5) node[midway, left] {$T_{85}$};
\draw[<-,  blue, dashed, line width=1pt] (5.1,3.0) -- (5.1,4.5) node[midway, left] {$T_{86}$};
\draw[<-,  green, dashed, line width=1pt] (5.6,2.0) -- (5.6,4.0) node[midway, left] {$T_{74}$};
\draw[<-, green, dashed, line width=1pt] (5.8,2.5) -- (5.8,4.0) node[midway, left] {$T_{75}$};
\draw[<-, green, dashed, line width=1pt] (6.0,3.0) -- (6.0,4.0) node[midway, left] {$T_{76}$};
  \node[above] at (1.5, 5) {$|1\rangle \leftrightarrow |{3}\rangle$};
  \node[above] at (5.0, 5) {$|2\rangle \leftrightarrow |{3}\rangle$};
\node[rotate=90, text width=2.0cm] at (-.5,0.7) {$\mathcal{E}(m,n)$};
\node[rotate=90, text width=2.5cm] at (-.5,2.5) {$\mathcal{E}(m-1,n+1)$};
\node[rotate=90, text width=2.5cm] at (-.5,4.5) {$\mathcal{E}(m-1,n)$};
\end{tikzpicture}
\caption*{Figure 8: Three energy manifolds of the dressed states of the $\Lambda$ configuration with all allowed transitions. For transition pathway $|3\rangle \rightarrow |{1}\rangle$, (i.e., from  energy manifold $\mathcal{E}(m-1,n)$ to $\mathcal{E}(m,n)$), the nonvanishing transition amplitude  is proportional to the coupling parameter $g_{13}$, while for the route $|2\rangle \rightarrow |{1}\rangle$, (i.e., from the energy manifold $\mathcal{E}(m-1,n)$ to $\mathcal{E}(m-1,n+1)$), the amplitude is proportional to $g_{23}$.}
\end{figure}
\par 
\begin{table}[ht]
 \caption{Allowed transitions in pathway $\ket{3} \rightarrow \ket{1}$}
  \centering
  \begin{tabular}{lll}
    \toprule
    \cmidrule(r){1-2}
    Sl. No.  & \qquad Transition in dressed states & Order of Peak  \\
    \midrule
    Transition I & $\ket{D(9,m-1,n)}$ $\Longrightarrow$ $\ket{D(1,m,n)}$ & 2nd oder \\
    \midrule   
    Transition II & $\ket{D(7,m-1,n)}$ $\Longrightarrow$ $\ket{D(2,m,n)}$ & 1st  order \\ 
      & $\ket{D(8,m,n)}$ $\Longrightarrow$ $\ket{D(3,m,n)}$\\
      \midrule
    Transition III & $\ket{D(7,m-1,n)}$ $\Longrightarrow$ $\ket{D(1,m,n)}$ & (Central Peak) \\ 
     & $\ket{D(8,m-1,n)}$ $\Longrightarrow$ $\ket{D(2,m,n)}$  & Zero-th order \\
     & $\ket{D(9,m-1,n)}$ $\Longrightarrow$ $\ket{D(3,m,n)}$ \\
     \midrule
     Transition IV & $\ket{D(9,m-1,n)}$ $\Longrightarrow$ $\ket{D(2,m,n)}$ & 1st order \\
      & $\ket{D(8,m-1,n)}$ $\Longrightarrow$ $\ket{D(1,m,n)}$\\
    \midrule 
     Transition V  & $\ket{D(7,m-1,n)}$ $\Longrightarrow$ $\ket{D(3,m,n)}$ & 2nd order \\
    \bottomrule
  \end{tabular}
  \label{tab:table}
\end{table}

\par
Finally for completeness, we discuss the dressed states of the three-level configuration which is indispensable part to understand the origin of quintuplet fluorescence spectrum and other related phenomena. It is well known that in presence of strong laser field, the dressed states $\ket{D(i,m,n)}$ is a coherent superposition of three atom-field bare states, namely,  
\begin{eqnarray}
\{\ket{-,m,n}, \quad \ket{0,m-1,n+1} \quad \ket{+,m-1,n}\},
\end{eqnarray}
where $\{\ket{-}, \ket{0}, \ket{+}\}$ represent the atomic states with $m$, $n$ be the photon numbers of the bi-chromatic fields. The first triplet of the dressed states for the $\Lambda$ configuration, which constitutes the energy manifold $\mathcal{E}({m,n})$, is given by \cite{Sen2012},
\begin{eqnarray} 
     \begin{bmatrix} \ket{D(1,m,n)} \\ \ket{D(2,m,n)} \\ \ket{D(3,m,n)} \end{bmatrix} =
     \mathcal{R}_{m,n}
     \begin{bmatrix} \ket{+,m-1,n} \\ \ket{0,m-1,n+1} \\ \ket{-,m,n} \end{bmatrix}.\;
\end{eqnarray}
where $\mathcal{R}_{m,n}(\theta,\phi,\psi)$ be the Euler matrix with mixing angles $\{\theta,\phi,\psi\}$ which mixes the bare states. This triplet provides the basis of constructing the tower of remaining dressed states with judicious choice of the photon numbers of the bare states. Fig.8 shows the first three manifolds which precisely preserves the topology of the $\Lambda$ configuration having two well-defined transition pathways, namely, $\ket{3} \rightarrow \ket{1}$ (solid lines) and $\ket{3}\rightarrow \ket{2}$ (dotted lines). The dipole transition amplitudes between $j$ to $i$-th dressed states of either of these pathways are given by the dipole matrix element, 
\begin{eqnarray}
d_{ij}=\bra{D(j,m,n)}\hat{\mathcal{O}}\ket{D(i,m,n)}, 
\end{eqnarray}
where $\hat{\mathcal{O}}$ be the interaction operator described by the Hamiltonian (\ref{eq3a}) proportional to the dipole coupling strength $g_{13}$ or $g_{23}$ depending upon the route. The exact evaluation of the transition amplitudes requires the knowledge of the Euler angles for different configurations which is beyond the scope of the paper.
\par 
To understand the quintuplet structure of the $\Lambda$ configuration of different order, in Table-I we have displayed all possible spontaneous transitions from the dressed states of manifold $\mathcal{E}(m-1,n)$ to those in $\mathcal{E}(m,n)$ at zero detuning, namely $\tilde{\Delta}_{23}=0$ and $\tilde{\Delta}_{13}=0$.  
It is worth noting here that corresponding to zero-th order central peak we have three transitions, then a pair of of sidebands with diminishing intensity with two and then single transitions, respectively. Similarly we have spontaneous transitions from the energy manifold $\mathcal{E}(m-1,n)$ to $\mathcal{E}(m-1, n+1)$ which has same structure (Table not shown). The treatment for the $V$ and $\Xi$ configurations of the three-level system is similar.

\section{Conclusion}
\par
The quantum interference in the $\Lambda$, $V$ and $\Xi$ - type of three-level configuration in presence of intense bi-chromatic resonant laser field exhibits several non-trivial features of this strong coupling phenomena. In this paper we have developed a comprehensive method to study the resonance fluorescence spectrum of  different three-level configurations. In contrast to other available methods \cite{Narducci1990,Fu1992,Ferguson1996}, our approach rooted in the SU(3) based group theoretical technique which offers a succinct derivation of the optical Bloch equation, presented in a concise format. When combined with the quantum regression theorem, this derivation enables the calculation of the fluorescence profiles of all configurations across different parameter regimes. Finally to explain the origin of quintuplet spectra, we have touched upon the phenomenological description of the dressed state of the $\Lambda$ system. The emergence of the heralded photons steaming from the transitions of the dressed states of various three-level configurations is another important aspect which requires further exploration of the structural aspects of the dressed states. In the parlance of significant development of the atom-field interaction, further study of the dressed states of the three-level system may unravel uncharted phenomena of quantum information processing. 

\pagebreak 

\appendix
\section{Bloch matrices of Lambda, Vee and Cascade configurations}
\par
In Equation (\ref{eq16}), the Bloch matrix $M^{\Lambda}$ of the $\Lambda$ configuration is found to be,\\
$M^{\Lambda}=$
\begin{equation}
\begin{pmatrix}
D_{11}^{\Lambda } & 0 & -\frac{2 i g_{23}}{3} & 0 & 0 & -\frac{i g_{23}}{3} & 0 & i g_{13} & \frac{i g_{23}}{3} \\
 0 & D_{22}^{\Lambda } & \frac{2 i g_{23}}{3} & 0 & 0 & \frac{i g_{23}}{3} & -i g_{13} & 0 & -\frac{i g_{23}}{3} \\
 -2 i g_{23} & 2 i g_{23} & D_{33}^{\Lambda } & -i g_{13} & i g_{13} & -\frac{\Gamma _{31}}{3} & 0 & 0 & \frac{1}{3} \left(2 \Gamma _{32}-\Gamma
   _{31}\right) \\
 0 & 0 & -\frac{i g_{13}}{3} & D_{44}^{\Lambda } & 0 & -\frac{2 i g_{13}}{3} & i g_{23} & 0 & -\frac{i g_{13}}{3} \\
 0 & 0 & \frac{i g_{13}}{3} & 0 & D_{55}^{\Lambda } & \frac{2 i g_{13}}{3} & 0 & -i g_{23} & \frac{i g_{13}}{3} \\
 -i g_{23} & i g_{23} & -\frac{\Gamma _{32}}{3} & -2 i g_{13} & 2 i g_{13} & D_{66}^{\Lambda } & 0 & 0 & \frac{1}{3} \left(\Gamma _{32}-2 \Gamma
   _{31}\right) \\
 0 & -i g_{13} & 0 & i g_{23} & 0 & 0 & D_{77}^{\Lambda } & 0 & 0 \\
 i g_{13} & 0 & 0 & 0 & -i g_{23} & 0 & 0 & D_{88}^{\Lambda } & 0 \\
 i g_{23} & -i g_{23} & \frac{\Gamma _{32}}{3} & -i g_{13} & i g_{13} & -\frac{\Gamma _{31}}{3} & 0 & 0 & D_{99}^{\Lambda }
\end{pmatrix} 
\end{equation}
where the diagonal terms are given by,
\begin{eqnarray}
    D_{11}^{\Lambda }&=&-\frac{\Gamma _{32}}{2}+i \Delta _{23}, \quad D_{22}^{\Lambda }=-\frac{\Gamma _{32}}{2}-i \Delta _{23},\quad D_{33}^{\Lambda}=-\frac{2 \Gamma _{32}}{3}, \nonumber\\    D_{44}^{\Lambda } &=&-\frac{\Gamma _{31}}{2}+i \Delta _{13},\quad D_{55}^{\Lambda}=-\frac{\Gamma _{31}}{2}-i \Delta _{13}, \quad D_{66}^{\Lambda }=-\frac{2 \Gamma _{31}}{3}\nonumber\\
   D_{77}^{\Lambda}&=&\frac{1}{2} \left(-\Gamma _{31}-\Gamma _{32}+2 i \left(\Delta _{13}-\Delta _{23}\right)\right),\\
   D_{88}^{\Lambda}&=& \frac{1}{2} \left(-\Gamma _{31}-\Gamma _{32}-2 i
   \left(\Delta _{13}-\Delta _{23}\right)\right), \quad
   D_{99}^{\Lambda}= \frac{1}{3} \left(-\Gamma _{31}-\Gamma _{32}\right) \nonumber
\end{eqnarray}

with the inhomogeneous term, 
\begin{equation}
\mathbf{B}^{\Lambda}=\big[0,0,\frac{1}{3} (\Gamma _{31}+2\Gamma _{32}),0, 0,\frac{1}{3} (2\Gamma _{31}+\Gamma _{32}), 0, 0, \frac{1}{3} (\Gamma _{31}-\Gamma_{32})\big]^T. 
\end{equation} 
\par 
For the $V$ configuration, the components of the density matrix in terms of Bloch vectors are given by,
\begin{eqnarray}\label{A.4}
\eqalign{\rho_{33}^V&=\frac{1}{3}(1+S_{T_{3}}^V+S_{V_{3}}^V), \quad \rho_{32}^V=S_{T_{-}}^V, \quad \rho_{31}^V=S_{V_{-}}^V, \\
\rho_{23}^V&=S_{T_{+}}^V, \quad \rho_{22}^V=\frac{1}{3}(1-S_{T_{3}}^V+S_{U_{3}}^V),
\quad \rho_{21}=S_{U_{-}}^V, \\
\rho_{13}^V&=S_{V_{+}}^V, \quad \rho_{12}^V=S_{U_{+}}^V, \quad \rho_{11}^V=\frac{1}{3}(1-S_{U_{3}}^V-S_{V_{3}}^V ).} \end{eqnarray}
Substituting Equation (\ref{A.4}) into (\ref{eq13b}), the OBE of the $V$ configuration is given by,
\begin{eqnarray}
\frac{d\mathbf{S^{V}_{\mathbb{P}_i}}(t)}{dt}= M^{V}\mathbf{S^{V}_{\mathbb{P}_i}}+\mathbf{B^{V}},
\end{eqnarray}
where the Bloch matrix $M^V$ is given by,\\
$M^V=$
\begin{equation}
\begin{pmatrix}
D_{11}^V & 0 & 0 & -i g_{12} & 0 & 0 & 0 & i g_{13} & 0 \\
 0 & D_{22}^V & 0 & 0 & i g_{12} & 0 & -i g_{13} & 0 & 0 \\
 0 & 0 & D_{33}^V & -i g_{13} & i g_{13} & -\frac{\Gamma _{31}}{3} & i g_{12} & -i g_{12} & \frac{\Gamma _{21}}{3} \\
 -i g_{12} & 0 & -\frac{i g_{13}}{3} & D_{44}^V & 0 & -\frac{2 i g_{13}}{3} & 0 & 0 & -\frac{i g_{13}}{3} \\
 0 & i g_{12} & \frac{i g_{13}}{3} & 0 & D_{55}^V & \frac{2 i g_{13}}{3} & 0 & 0 & \frac{i g_{13}}{3} \\
 0 & 0 & \frac{1}{3} \left(\Gamma _{21}-2 \Gamma _{31}\right) & -2 i g_{13} & 2 i g_{13} & D_{66}^V & -i g_{12} & i g_{12} & -\frac{\Gamma
   _{21}}{3} \\
 0 & -i g_{13} & \frac{i g_{12}}{3} & 0 & 0 & -\frac{i g_{12}}{3} & D_{77}^V & 0 & -\frac{2 i g_{12}}{3} \\
 i g_{13} & 0 & -\frac{i g_{12}}{3} & 0 & 0 & \frac{i g_{12}}{3} & 0 & D_{88}^V & \frac{2 i g_{12}}{3} \\
 0 & 0 & \frac{1}{3} \left(2 \Gamma _{21}-\Gamma _{31}\right) & -i g_{13} & i g_{13} & -\frac{\Gamma _{31}}{3} & -2 i g_{12} & 2 i g_{12} &
   D_{99}^V 
\end{pmatrix} 
\end{equation}
where the diagonal terms are given by
\begin{eqnarray}
    D_{11}^V &=&\frac{1}{2} \left(-\Gamma _{21}-\Gamma _{31}-2 i \left(\Delta _{12}-\Delta _{13}\right)\right),\nonumber \\ D_{22}^V&=&\frac{1}{2} \left(-\Gamma _{21}-\Gamma _{31}+2 i \left(\Delta _{12}-\Delta _{13}\right)\right), \quad D_{33}=\frac{1}{3} \left(-\Gamma _{21}-\Gamma _{31}\right), \nonumber \\ 
   D_{44}^V&=&-\frac{\Gamma _{31}}{2}+i \Delta _{13}, \quad   D_{55}^V =-\frac{\Gamma _{31}}{2}-i \Delta _{13}, \quad
   D_{66}^V=-\frac{2 \Gamma _{31}}{3},  \\
   \quad D_{77}^V &=&-\frac{\Gamma _{21}}{2}+i \Delta _{12}, \quad  
   D_{88}^V =-\frac{\Gamma _{21}}{2}-i \Delta _{12}, \quad D_{99}^V=-\frac{2 \Gamma _{21}}{3} \nonumber
\end{eqnarray}
with the corresponding inhomogeneous term,
\begin{equation}
B^{V}=\big[0, 0,\frac{1}{3} (\Gamma _{21}-\Gamma _{31}), 0, 0,-\frac{1}{3} (\Gamma _{21}+2\Gamma_{31}),0,0,-\frac{1}{3}(2\Gamma_{21}+\Gamma _{31})\big]^T, 
\end{equation}
\par
Finally for the $\Xi$ configuration, the components of the density matrix in terms of Bloch vectors are given by,
\begin{eqnarray}\label{A.9}
\eqalign{\rho_{33}^\Xi&=\frac{1}{3}(1+S_{T_{3}}^\Xi+S_{V_{3}}^\Xi), \quad \rho_{32}^\Xi=S_{T_{-}}^\Xi, \quad \rho_{31}^\Xi=S_{V_{-}}^\Xi, \\
\rho_{23}^\Xi&=S_{T_{+}}^\Xi, \quad \rho_{22}^\Xi=\frac{1}{3}(1-S_{T_{3}}^\Xi+S_{U_{3}}^\Xi),
\quad \rho_{21}^\Xi=S_{U_{-}}^\Xi, \\
\rho_{13}^\Xi&=S_{V_{+}}^\Xi, \quad \rho_{12}^\Xi=S_{U_{+}}^\Xi, \quad \rho_{11}^\Xi=\frac{1}{3}(1-S_{U_{3}}^\Xi-S_{V_{3}}^\Xi).}
\end{eqnarray}
Substituting (\ref{A.9}) into (\ref{eq13c}) the OBE of the $\Xi$ configuration is given by,
\begin{eqnarray}
\frac{d\mathbf{S^{\Xi}_{\mathbb{P}_i}}(t)}{dt}= M^{\Xi}\mathbf{S^{\Xi}_{\mathbb{P}_i}}+\mathbf{B^{\Xi}},
\end{eqnarray}
with the Bloch matrix given by,
$M^{\Xi}=$
\begin{equation}
\begin{pmatrix}
 D_{11}^{\Xi } & 0 & -\frac{2 i g_{23}}{3} & -i g_{12} & 0 & -\frac{i g_{23}}{3} & 0 & 0 & \frac{i g_{23}}{3} \\
 0 & D_{22}^{\Xi } & \frac{2 i g_{23}}{3} & 0 & i g_{12} & \frac{i g_{23}}{3} & 0 & 0 & -\frac{i g_{23}}{3} \\
 -2 i g_{23} & 2 i g_{23} & D_{33}^{\Xi } & 0 & 0 & -\frac{2 \Gamma _{32}}{3} & i g_{12} & -i g_{12} & \frac{\Gamma _{21}}{3} \\
 -i g_{12} & 0 & 0 & D_{44}^{\Xi } & 0 & 0 & i g_{23} & 0 & 0 \\
 0 & i g_{12} & 0 & 0 & D_{55}^{\Xi } & 0 & 0 & -i g_{23} & 0 \\
 -i g_{23} & i g_{23} & \frac{1}{3} \left(\Gamma _{21}-\Gamma _{32}\right) & 0 & 0 & D_{66}^{\Xi } & -i g_{12} & i g_{12} & -\frac{\Gamma
   _{21}}{3} \\
 0 & 0 & \frac{i g_{12}}{3} & i g_{23} & 0 & -\frac{i g_{12}}{3} & D_{77}^{\Xi } & 0 & -\frac{2 i g_{12}}{3} \\
 0 & 0 & -\frac{i g_{12}}{3} & 0 & -i g_{23} & \frac{i g_{12}}{3} & 0 & D_{88}^{\Xi } & \frac{2 i g_{12}}{3} \\
 i g_{23} & -i g_{23} & \frac{1}{3} \left(2 \Gamma _{21}+\Gamma _{32}\right) & 0 & 0 & \frac{\Gamma _{32}}{3} & -2 i g_{12} & 2 i g_{12} &
   D_{99}^{\Xi } 
\end{pmatrix} 
\end{equation}
where the diagonal terms are given by 
\fl \begin{eqnarray}
D_{11}^{\Xi}&=& \frac{1}{2} \left(-\Gamma _{21}-\Gamma _{32}+2 i \Delta _{23}\right), \quad D_{22}^{\Xi }=\frac{1}{2} \left(-\Gamma _{21}-\Gamma _{32}-2 i \Delta _{23}\right),\nonumber\\ D_{33}^{\Xi }&=&\frac{1}{3} \left(-\Gamma _{21}-2 \Gamma _{32}\right), \quad  
   D_{44}^{\Xi }=-\frac{\Gamma _{32}}{2}+i \left(\Delta _{12}+\Delta _{23}\right), \nonumber\\  D_{55}^{\Xi}&=&-\frac{\Gamma _{32}}{2}-i \left(\Delta _{12}+\Delta _{23}\right),\quad D_{66}^{\Xi }=-\frac{\Gamma _{32}}{3}, \\   D_{77}^{\Xi }&=&-\frac{\text{$\Gamma $21}}{2}+i \Delta _{12}, \quad 
   D_{88}^{\Xi}=-\frac{\Gamma _{21}}{2}-i \Delta _{12}, \quad
  D_{99}^{\Xi }=-\frac{2 \Gamma _{21}}{3}\nonumber
\end{eqnarray}
with corresponding inhomogeneous term,
\begin{equation}
\mathbf{B}^{\Xi}=\big[0, 0,\frac{1}{3}(\Gamma _{21}-2\Gamma _{32}), 0,0,-\frac{1}{3} (\Gamma _{21}+\Gamma _{32}), 0, 0,\frac{1}{3}(\Gamma _{32}-2\Gamma_{21})\big]^T. 
\end{equation}

\newpage 

\printbibliography

\newpage

\section*{Figure captions:}
\begin{figure}[ht]
   \centering
    \includegraphics[scale=0.7]{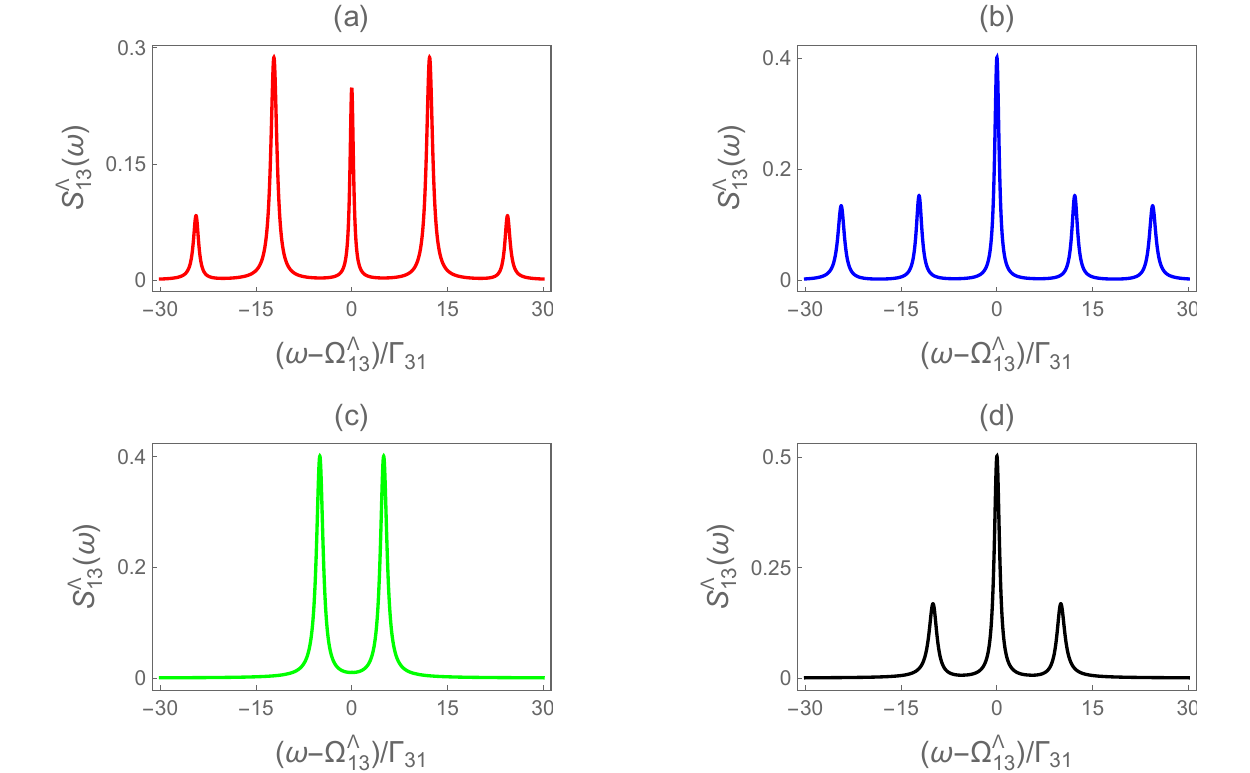}
    \caption*{Figure 2: Power spectrum of $\Lambda$ configuration for $\ket{3} \rightarrow \ket{1}$ ($\langle\langle  V_+(\tau)V_-(0) \rangle \rangle$ correlation) transition (Scaled with $\Gamma_{31}^\Lambda$): Plot of $S_{13}^{\Lambda}(\omega)$ versus $(\omega - \Omega_{13}^\Lambda)/\Gamma_{31}^\Lambda$ with following parameters: a) $\tilde{\Gamma}_{32}=.5$, $\tilde{g}_{23}=10$, $\tilde{g}_{13}=7$ (Red),
    b) $\tilde{\Gamma}_{32}=.5$, $\tilde{g}_{23}=7$, $\tilde{g}_{13}=10$ (Blue), 
    c) $\tilde{\Gamma}_{32}=.5$, $\tilde{g}_{23}=5$, $\tilde{g}_{13}=.1$ (Green),
    d) $\tilde{\Gamma}_{32}=.5$, $\tilde{g}_{23}=.1$, $\tilde{g}_{13}=5$ (Black)}
    \label{Fig:fig2}
\end{figure}
\begin{figure}[ht]
   \centering
    \includegraphics[scale=0.7]{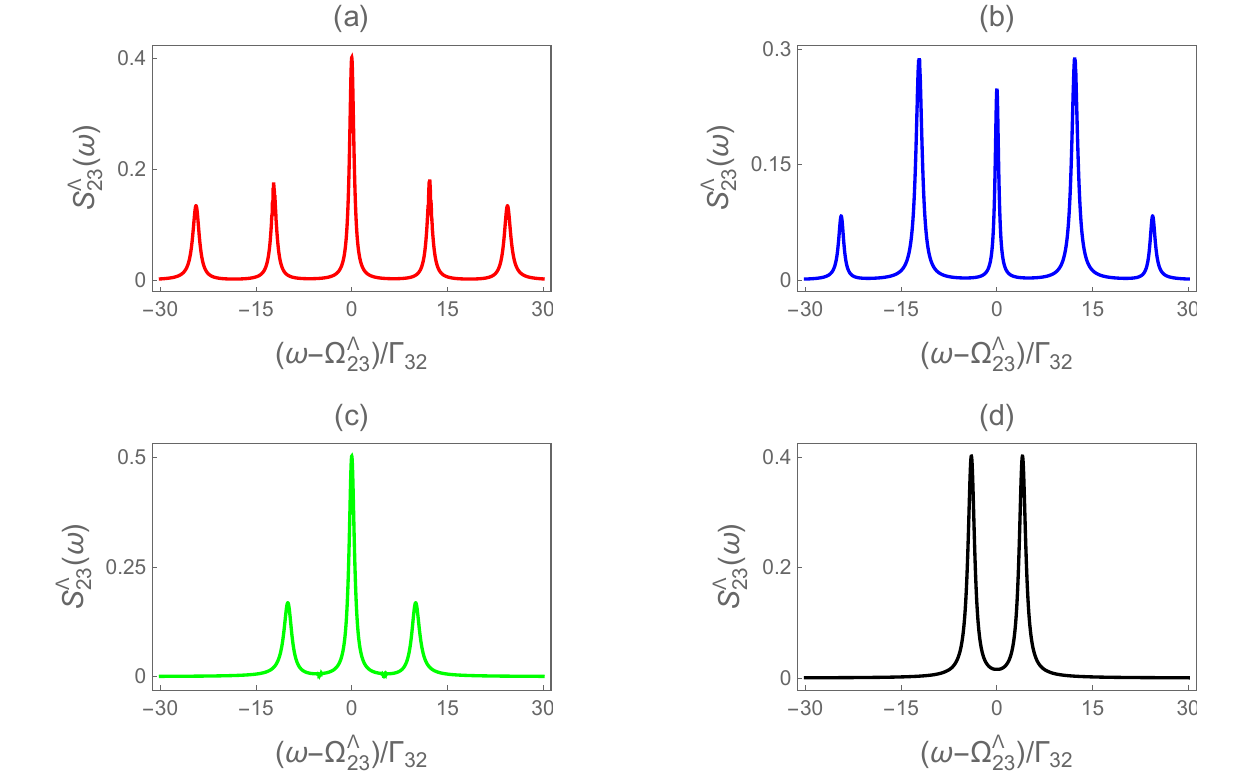}
    \caption*{Figure 3: Power spectrum of $\Lambda$ configuration for $\ket{3} \rightarrow \ket{2}$ ($\langle\langle T_+(\tau)T_-(0) \rangle \rangle$ correlation) transition (Scaled with $\Gamma_{32}^\Lambda$): Plot of $S_{13}^{\Lambda}(\omega)$ versus $(\omega - \Omega_{13}^\Lambda)/\Gamma_{32}^\Lambda$ with following parameters: 
    a) $\tilde{\Gamma}_{31}=.5$, $\tilde{g}_{23}=10$, $\tilde{g}_{13}=7$ (Red),
    b) $\tilde{\Gamma}_{31}=.5$, $\tilde{g}_{23}=7$, $\tilde{g}_{13}=10$ (Blue), 
    c) $\tilde{\Gamma}_{31}=.5$, $\tilde{g}_{23}=5$, $\tilde{g}_{13}=.1$ (Green),
    d) $\tilde{\Gamma}_{31}=.5$, $\tilde{g}_{23}=.1$, $\tilde{g}_{13}=5$ (Black)}
    \label{fig3}
\end{figure}

\newpage

\begin{figure}[ht]
   \centering
    \includegraphics[scale=0.7]{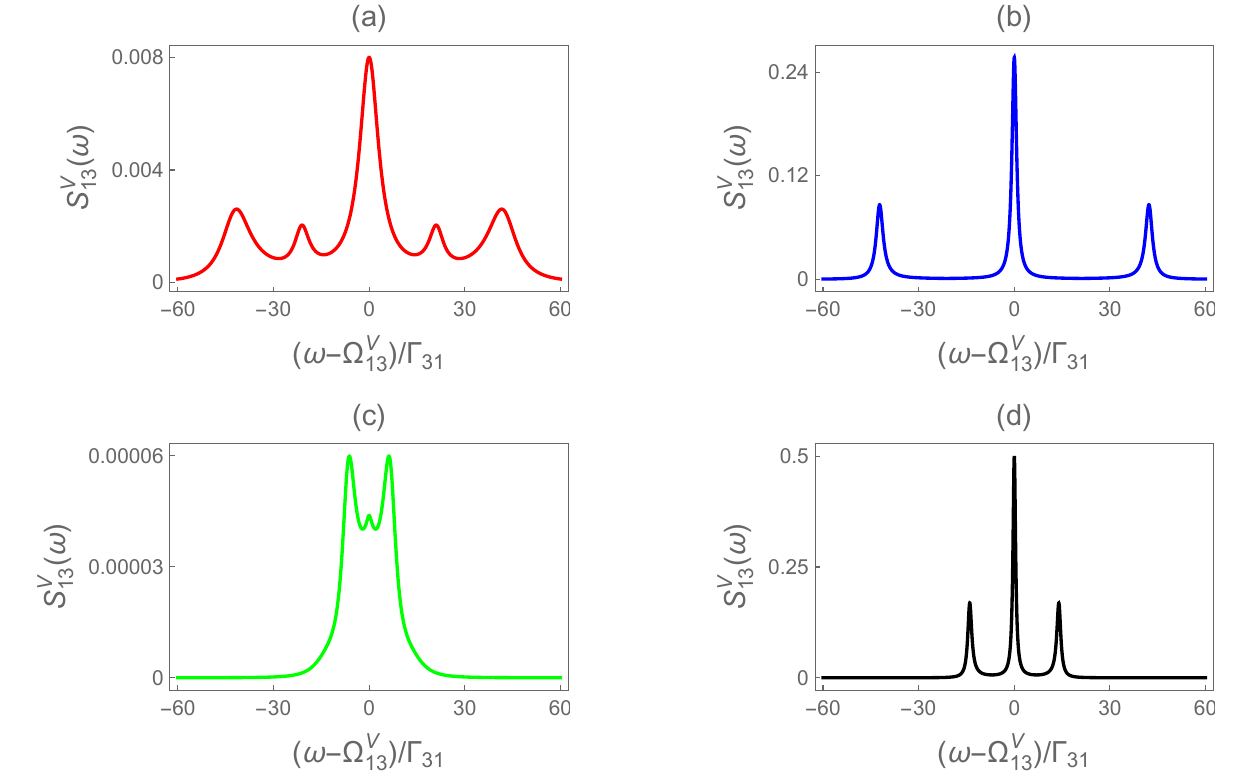}
    \caption*{Figure 4: Power spectrum of $V$ configuration for $\ket{3} \rightarrow \ket{1}$ ($\langle\langle V_+(\tau)V_-(0) \rangle \rangle$ correlation) transition (Scaled with $\Gamma_{31}^V$): Plot of $S_{13}^{V}(\omega)$ versus $(\omega - \Omega_{13}^V)/\Gamma_{31}^V$ with following parameters: 
    a) $\tilde{\Gamma}_{21}=8$, $\tilde{g}_{12}=20$, $\tilde{g}_{13}=7$ (Red),
    b) $\tilde{\Gamma}_{21}=8$, $\tilde{g}_{12}=7$, $\tilde{g}_{13}=20$ (Blue), 
    c) $\tilde{\Gamma}_{21}=8$, $\tilde{g}_{12}=7$, $\tilde{g}_{13}=.1$ (Green),
    d) $\tilde{\Gamma}_{21}=8$, $\tilde{g}_{12}=.1$, $\tilde{g}_{13}=7$ (Black).
     }
    \label{fig4}
\end{figure}

\begin{figure}[ht]
   \centering
    \includegraphics[scale=0.7]{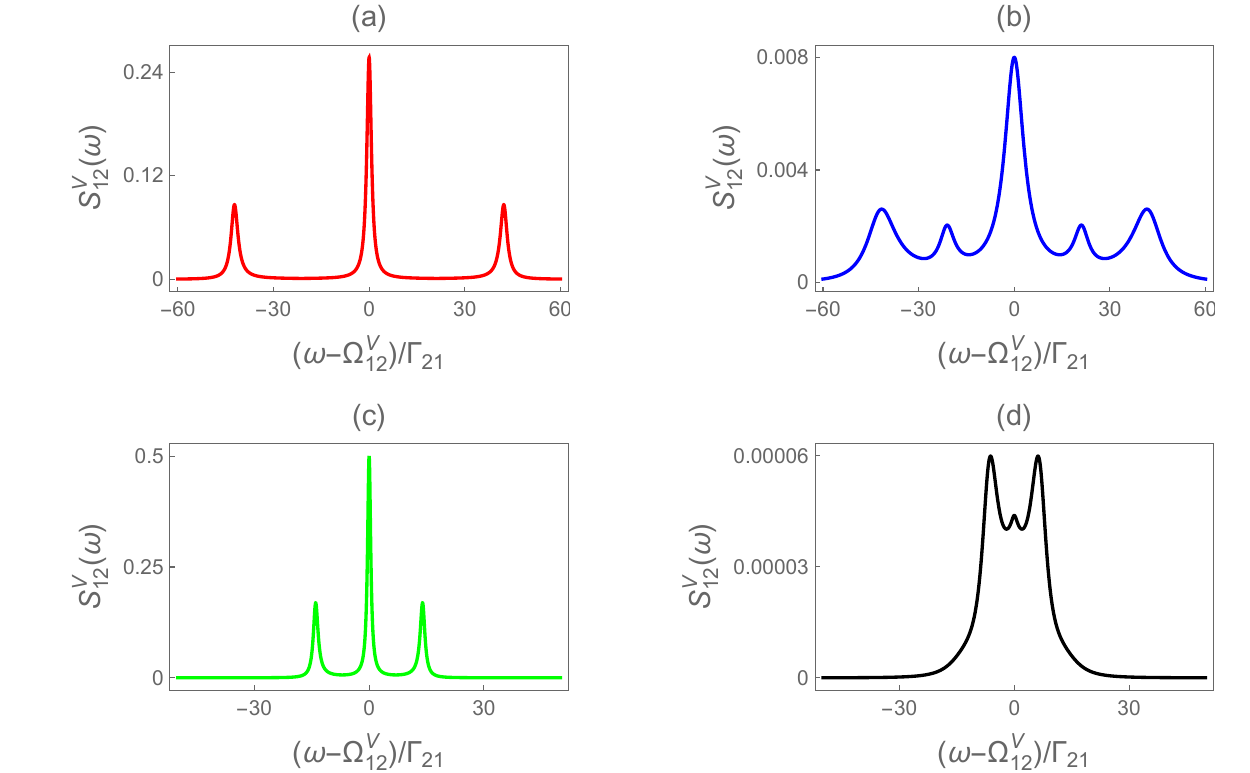}
    \caption*{Figure 5: Power spectrum of $V$ configuration for $\ket{2} \rightarrow \ket{1}$ ($\langle\langle U_+(\tau)U_-(0) \rangle \rangle$ correlation) transition (Scaled with $\Gamma_{21}^V$): Plot of $S_{12}^{V}(\omega)$ versus $(\omega - \Omega_{12}^V)/\Gamma_{21}^V$ with following parameters: 
     a) $\tilde{\Gamma}_{31}=8$, $\tilde{g}_{12}=20$, $\tilde{g}_{13}=7$ (Red),
    b) $\tilde{\Gamma}_{31}=8$, $\tilde{g}_{12}=7$, $\tilde{g}_{13}=20$ (Blue), 
    c) $\tilde{\Gamma}_{31}=8$, $\tilde{g}_{12}=7$, $\tilde{g}_{13}=.1$ (Green),
    d) $\tilde{\Gamma}_{31}=8$, $\tilde{g}_{12}=.1$, $\tilde{g}_{13}=7$ (Black)
     }
    \label{fig5}
\end{figure}

\pagebreak

\begin{figure}[h]
   \centering
    \includegraphics[scale=0.7]{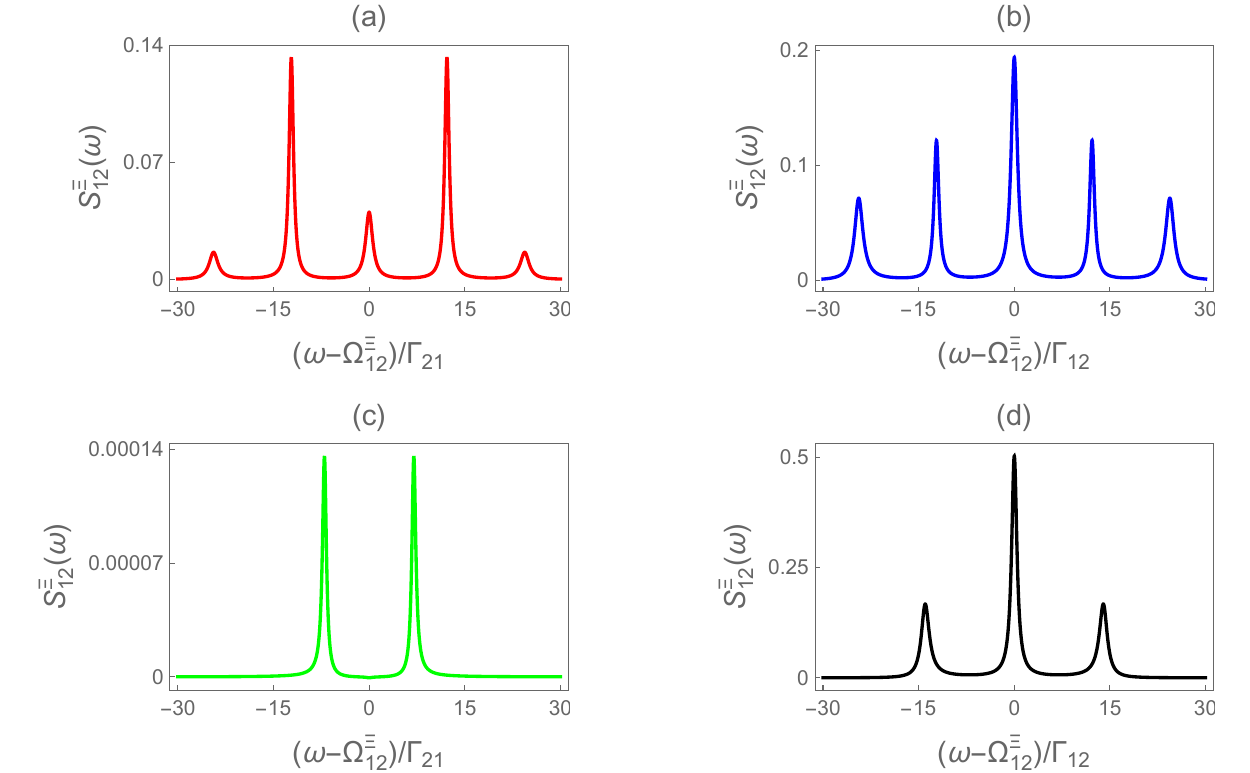}
    \caption*{Figure 6: Power spectrum of $\Xi$ configuration for $\ket{2} \rightarrow \ket{1}$ ($\langle\langle U_+(\tau)U_-(0) \rangle \rangle$ correlation) transition (Scaled with $\Gamma_{21}^\Xi$): Plot of $S_{12}^{\Xi}(\omega)$ versus $(\omega - \Omega_{12}^\Xi)/\Gamma_{21}^\Xi$ with following parameters: 
    a) $\tilde{\Gamma}_{32}=.5$, $\tilde{g}_{23}=10$, $\tilde{g}_{12}=7$ (Red),
    b) $\tilde{\Gamma}_{32}=.5$, $\tilde{g}_{23}=7$, $\tilde{g}_{12}=10$ (Blue), 
    c) $\tilde{\Gamma}_{32}=.5$, $\tilde{g}_{23}=7$, $\tilde{g}_{12}=.1$ (Green),
    d) $\tilde{\Gamma}_{32}=.5$, $\tilde{g}_{23}=.1$, $\tilde{g}_{12}=7$ (Black)}
    \label{fig6}
\end{figure}

\begin{figure}[h]
   \centering
    \includegraphics[scale=0.7]{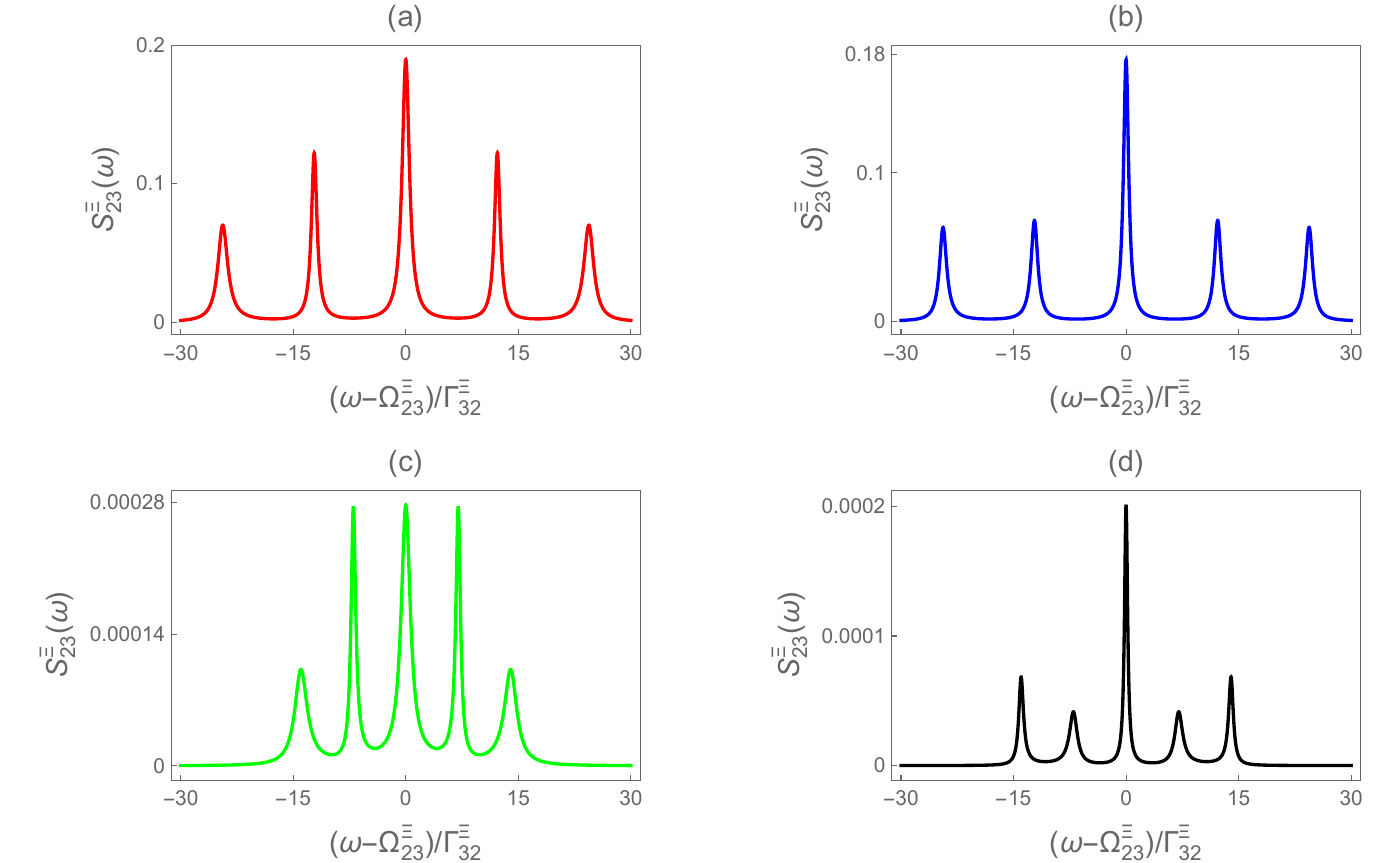}
    \caption*{Figure 7: Power spectrum of $\Xi$ configuration for $\ket{3} \rightarrow \ket{2}$ ($\langle\langle T_+(\tau)T_-(0) \rangle \rangle$ correlation) transition (Scaled with $\Gamma_{32}^\Xi$): Plot of $S_{23}^{\Xi}(\omega)$ versus $(\omega - \Omega_{23}^\Xi)/\Gamma_{32}^\Xi$ with following parameters: 
    a) $\tilde{\Gamma}_{21}=.5$, $\tilde{g}_{23}=10$, $\tilde{g}_{12}=7$ (Red),
    b) $\tilde{\Gamma}_{21}=.5$, $\tilde{g}_{23}=7$, $\tilde{g}_{12}=10$ (Blue), 
    c) $\tilde{\Gamma}_{21}=.5$, $\tilde{g}_{23}=7$, $\tilde{g}_{12}=.1$ (Green),
    d) $\tilde{\Gamma}_{21}=.5$, $\tilde{g}_{23}=.1$, $\tilde{g}_{12}=7$ (Black)}
    \label{fig7}
\end{figure}


\end{document}